\documentclass[printer]{aa}
\usepackage{graphicx}
\usepackage{txfonts}

\newlength{\bibitemsep}
\newlength{\bibparskip}
\let\oldthebibliography\thebibliography
\renewcommand\thebibliography[1]{%
  \oldthebibliography{#1}%
  \setlength{\parskip}{\bibitemsep}%
  \setlength{\itemsep}{\bibparskip}%
}

\begin{document} 

\title{A Photometric in-depth look at the core-collapsed globular cluster NGC 6284
           \thanks{Based on observations collected with the NASA/ESA HST (Prop. 15232, PI:Ferraro), obtained at the Space Telescope Science Institute, which is operated by AURA, Inc., under NASA contract NAS5-26555.}
          }

\author{Dan Deras \inst{1}
          \and
          Mario Cadelano \inst{1,2}
          \and 
          Barbara Lanzoni \inst{1,2}
          \and
          Francesco R. Ferraro \inst{1,2}
          \and
          Cristina Pallanca \inst{1,2}
          \and  
          Emanuele Dalessandro \inst{2} 
          \and    
          Alessio Mucciarelli \inst{1,2}      
}

\institute{
   Dipartimento di Fisica e Astronomia  ``Augusto Righi'', Alma Mater Studiorum, Universit\`{a}  di Bologna, Via Gobetti 93/2, I-40129 Bologna, Italy
   \and
   INAF - Osservatorio di Astrofisica e Scienza dello Spazio di Bologna, Via Gobetti 93/3, 40129 Bologna, Italy
}

\authorrunning{Deras et al.}
\titlerunning{Photometric in-depth look at NGC 6284}

\abstract{
High-resolution Hubble Space Telescope (\textit{HST}) optical observations have been
used to perform the deepest photometric study of the poorly studied
Galactic globular cluster NGC 6284. The deep colour-magnitude diagram
(CMD) that we obtained reaches 6 magnitudes below the main sequence
turn-off. We provide the first determination of the gravitational
centre ($C_{\rm grav}$) and density profile of the system from
resolved stars.  $C_{\rm grav}$ is significantly offset (by
$1.5-3\arcsec$) from the values in the literature.  The density
profile shows the presence of a steep central cusp, unambiguously
indicating that the cluster experienced the core-collapse phase.
Updated values of the structural parameters and relaxation times of
the system are provided.  We also constructed the first
high-resolution reddening map in the cluster direction, which allowed
us to correct the evolutionary sequences in the CMD for the effects of
differential reddening. Isochrone fitting to the corrected CMD
provided us with new estimates of the cluster age, average colour
excess, metallicity, and distance. We find an absolute age of $13.3
\pm 0.4$ Gyr, an average colour excess $E(B-V) = 0.32 \pm 0.01$, a
metallicity [Fe/H]$= -1.36 \pm 0.01$, and a true distance modulus
$(m-M)_0 = 15.61 \pm 0.04$ that sets the cluster distance at $13.2 \pm
0.2$ kpc from the Sun.  The superb quality of the CMD allowed a
clear-cut identification of the Red Giant Branch (RGB) bump, which is clearly
distinguishable along the narrow RGB. The absolute magnitude of this
feature turns out to be $\sim 0.2$ mag fainter than previous
identification.
}


\maketitle
%
\section{Introduction}
\label{sec:intro}
Globular clusters (GCs) are the oldest and largest stellar aggregates
in the Galaxy, with ages $t> 10$ Gyr and stellar populations made of
up to $10^6$ stars. They orbit the Galactic halo and bulge, and the
study of their structural, kinematic and dynamical properties allows
us to trace back the formation history of the Galaxy up to its very
early stages.
In this context, we are performing a detailed kinematic exploration
(see the results of the ESO Multi-Instrument Kinematic Survey - MIKiS
in \citealp{ferraro+18a, lanzoni+18a, lanzoni+18b, leanza+22,
  leanza+23, pallanca+23}) and photometric investigation of GCs in
different Galactic environments \citep[e.g.,][]{origlia+02,origlia+03,
  Valenti2010, dalessandro+14, Pallanca2019, Cadelano2020a, Raso2020,
  Pallanca2021a,Deras2023} with the aim to perform a complete
characterisation of these stellar systems in terms of structure, age,
stellar content (see examples in \citealt{ferraro+97, Ferraro2009,
  Lanzoni2010, Miocchi2013, dalessandro+13a,
  Pallanca2021b,Cadelano2022}, and dynamical evolutionary stage
\citep{ferraro+12, ferraro+18b, ferraro+19, ferraro+23,
  lanzoni+16}. This approach led to the discovery of a few anomalous
stellar systems in the Galactic bulge harbouring multi-age and
multi-iron sub-populations (see \citealp{Ferraro2009, Ferraro2016,
  Ferraro2021, Origlia2011, Massari2014, dalessandro+22, crociati+23,
  romano+23}), which might be remnants of giant clumps of gas and
stars that, 12 Gyr ago, contributed to generate the Galactic bulge.

\begin{figure*}[ht!]
\includegraphics[scale=0.40]{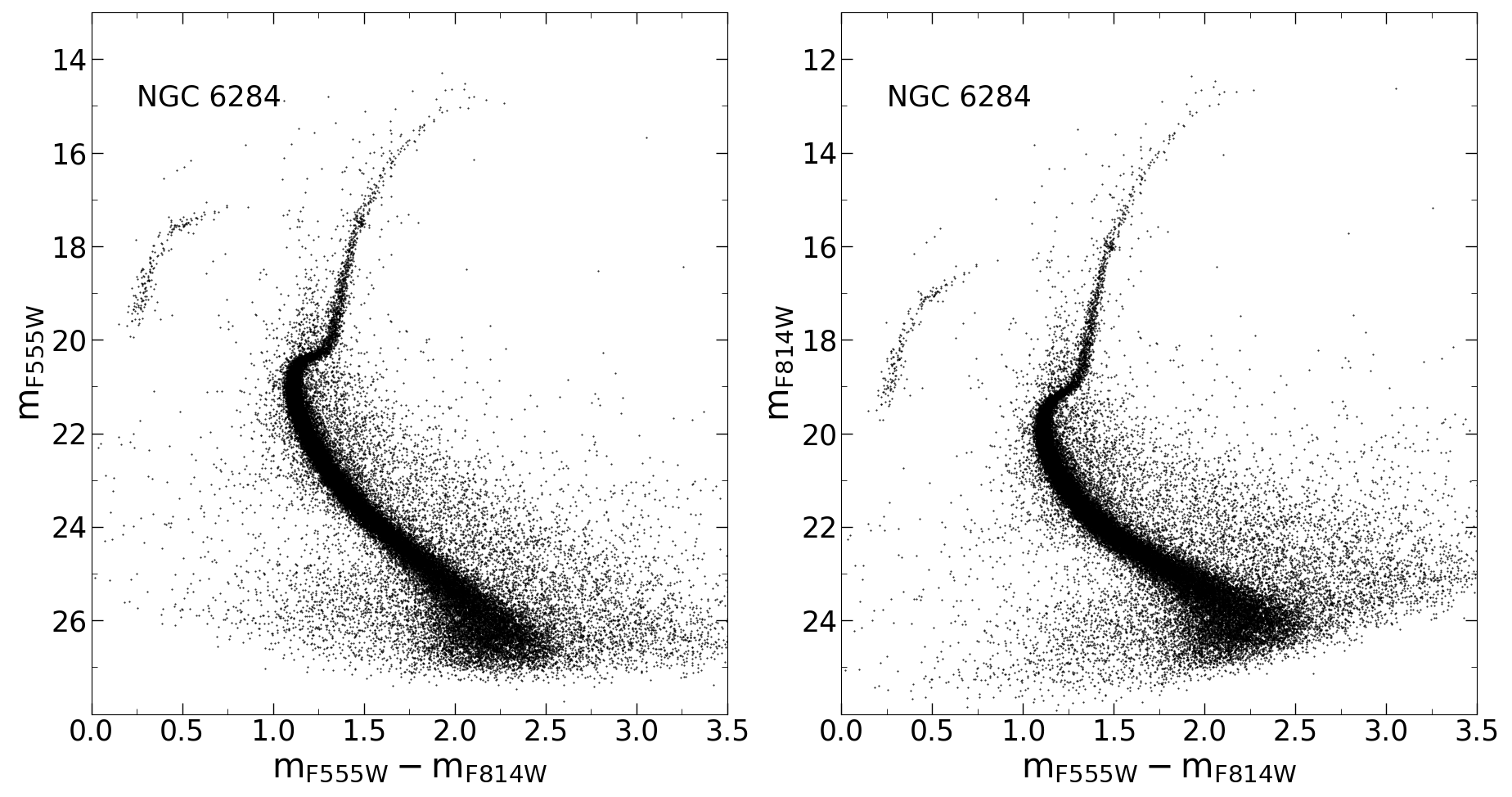}
\caption{CMD of NGC 6284 obtained from the $HST$ WFC3 data set used in
  this work. F555W and F814W magnitudes have been plotted along the
  y-axis on the left and right panels, respectively.}
\label{fig:cmd_complete}
\end{figure*}

The target of the present paper is the poorly studied Galactic GC NGC
6284. This is a cluster possibly associated to the thick disk (see
\citealt{casetti+10}) and located in the background of the Galactic
bulge: in projection in the sky it appears north of the bulge, at
Galactocentric coordinates $l=358.35\deg$ and $b=9.94\deg$.  According
to the \citet[][2010 edition]{Harris1996} catalogue, it is located at a
distance of 15.2 kpc from the Sun, while \citet{Baumgardt2021} reports
14.2 kpc. It follows a highly eccentric orbit
\citep[][$e=0.83$]{Bajkova2021} that extends up to $\sim$5.5 kpc along
the meridional direction of the Galactic plane, with a perigalactic
distance of 6.32 kpc \citep{Baumgardt2018}.  From the analysis of the
cluster integrals of motion, \citet{Massari2019} suggest that NGC 6284
may not have been generated in situ, but it has been rather accreted
by the Milky Way from the Gaia-Enceladus Sausage. On the other hand,
\citet{Bellazzini2020} ruled out NGC 6284 to be associated to the
Sagittarius dwarf spheroidal galaxy. In addition, based on the
destruction rates estimated by \citet{Gnedin1997}, \citet{Kundu2022}
suggest that the system is substantially affected by bulge and disk
shocks during its orbit crossing the Galactic plane.  From shape of
its surface brightness profile, NGC 6284 has been catalogued as a
post-core collapse (PCC) cluster since the very first surveys of
Galactic PCC systems (see \citealt{djo+86}; see also
\citealt{Trager1993, Trager1995, lugger+95}).  The PCC state is the
most advanced phase of internal dynamical evolution of collisional
systems. Recurrent stellar encounters cause kinetic-energy exchanges
among stars, with the most massive stars therefore transferring
kinetic energy to lower mass objects and progressively sinking toward
the system's centre. The energy transfer from the centre to the
cluster outer regions yields a progressive contraction of the core
leading to core collapse. The event is accompanied by a substantial
increase of the central density, and its observational manifestation
is the set up of a steep power-law cusp in the innermost region of the
projected density profile (\citealt{Meylan1997}, see also
\citealt{bhat+22}). Approximately 20$\%$ of the GCs in the Galaxy are
currently in this dynamical evolutionary stage \citep{Trager1993,
  Bianchini2018}.  In term of its chemical properties, there is a
general consensus on the fact that NGC 6284 is a mid-metallicity
cluster, but no results from high-resolution spectroscopy are
available at the moment, and the values reported in the literature
show a considerable spread: [Fe/H]$=-0.91$ \citep[][from RR
  Lyrae]{Smith1982}, [Fe/H]$=-0.99$ \citep[][from the measure of
  Calcium abudance]{Usher2019}, [Fe/H]$=-1.07$ \citep[][from
  low-resolution spectra of just 7 stars]{Dias2016}, and
[Fe/H]$=-1.26$ \citep{Harris1996}, which is the value obtained from
low-resolution spectra by \citet[][{[Fe/H]}$=-1.40$]{zinnwest84} after
conversion to the Carretta-Gratton scale.
The system is affected by a moderate reddening: $E(B-V)= 0.27-0.29$ in
most of the available studies \citep[e.g.,][]{Smith1982, Webbink1985,
  Zinn1985, Recio_Blanco2005}, although \citet{Minniti1995} quote
$E(B-V)= 0.40$.
Given its location toward the bulge direction, the extinction is
likely to show differential variations within cluster field of view.
Currently, there is no consensus about the absolute age of NGC
6284. The available estimates in the literature are: $t=11.0 \pm 0.1$
Gyr \citep{Deangeli2005}, $t=11.00 \pm 0.25$ Gyr \citep{Meissner2006},
$t=11.8 \pm 0.3$ Gyr \citep{Cabrera2022}, $t=12.3$ Gyr
\citep{Carretta2010}, and $t=13.0$ Gyr \citep{Dias2016}.

The goal of this work is to carry out the first in-depth photometric
analysis of NGC 6284 to characterise its structure and stellar
population through a set of high-resolution images sampling its
innermost regions.  The paper is structured as follows: In Section
\ref{sec:photometry} we describe the photometric analysis performed on
the acquired data-set. In Section \ref{sec:rad_prof} we determine the
centre of gravity of the cluster, its projected density profile and
its structural parameters. In Section \ref{sec:diff_red} we describe
the differential reddening correction performed on the
colour-magnitude diagram (CMD). In Section \ref{sec:age} we perform
isochrone fitting to the differential reddening corrected CMD to
estimate the age, distance, metallicity and absolute colour excess of
the system. Section \ref{sec:bump} is devoted to the determination of
the Red Giant Branch (RGB) bump luminosity from both the differential
and the integrated luminosity functions.  Finally, in Section
\ref{sec:summary} we summarise our results.

\begin{figure*}[ht!]
\begin{center}
\includegraphics[scale=0.21]{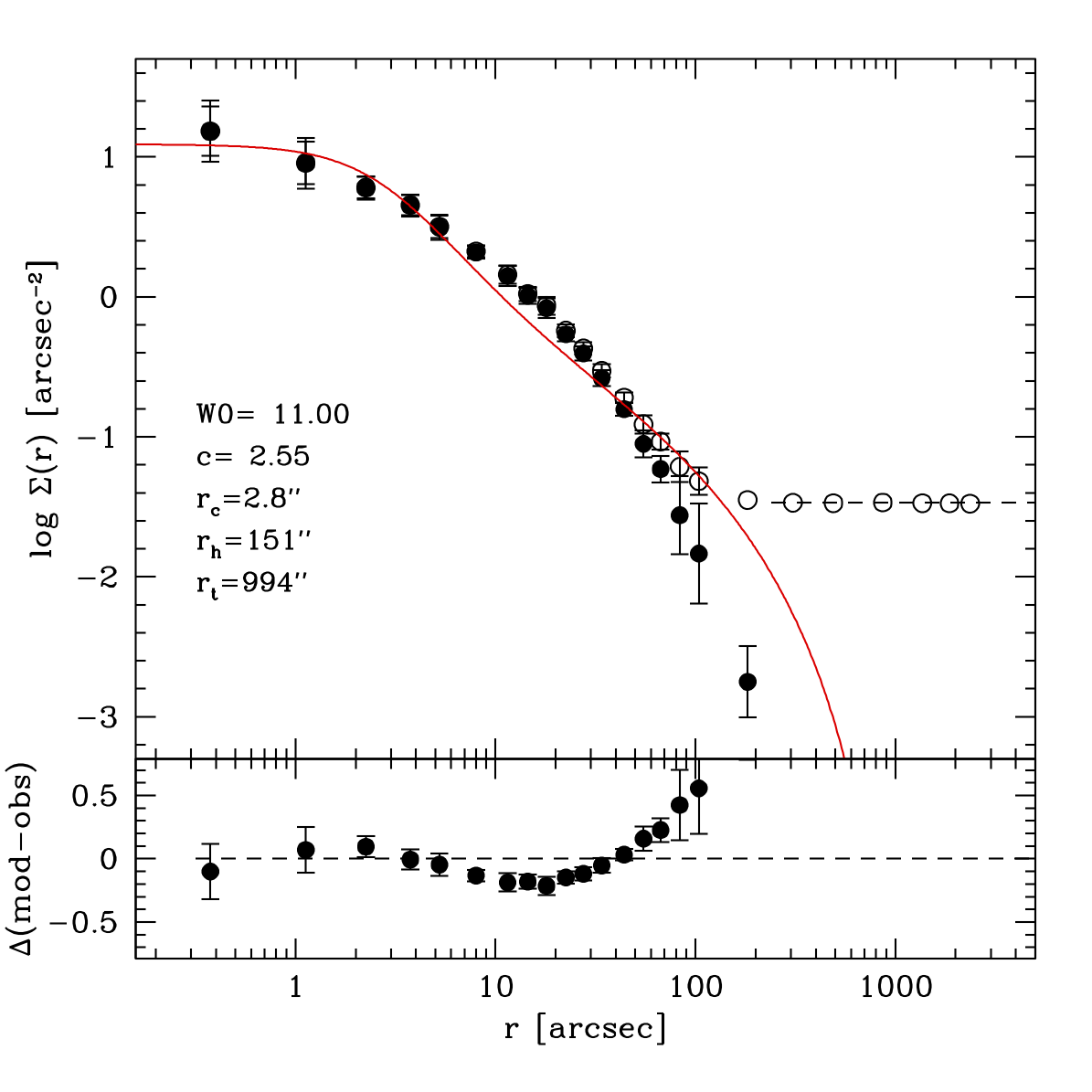}
\includegraphics[scale=0.21]{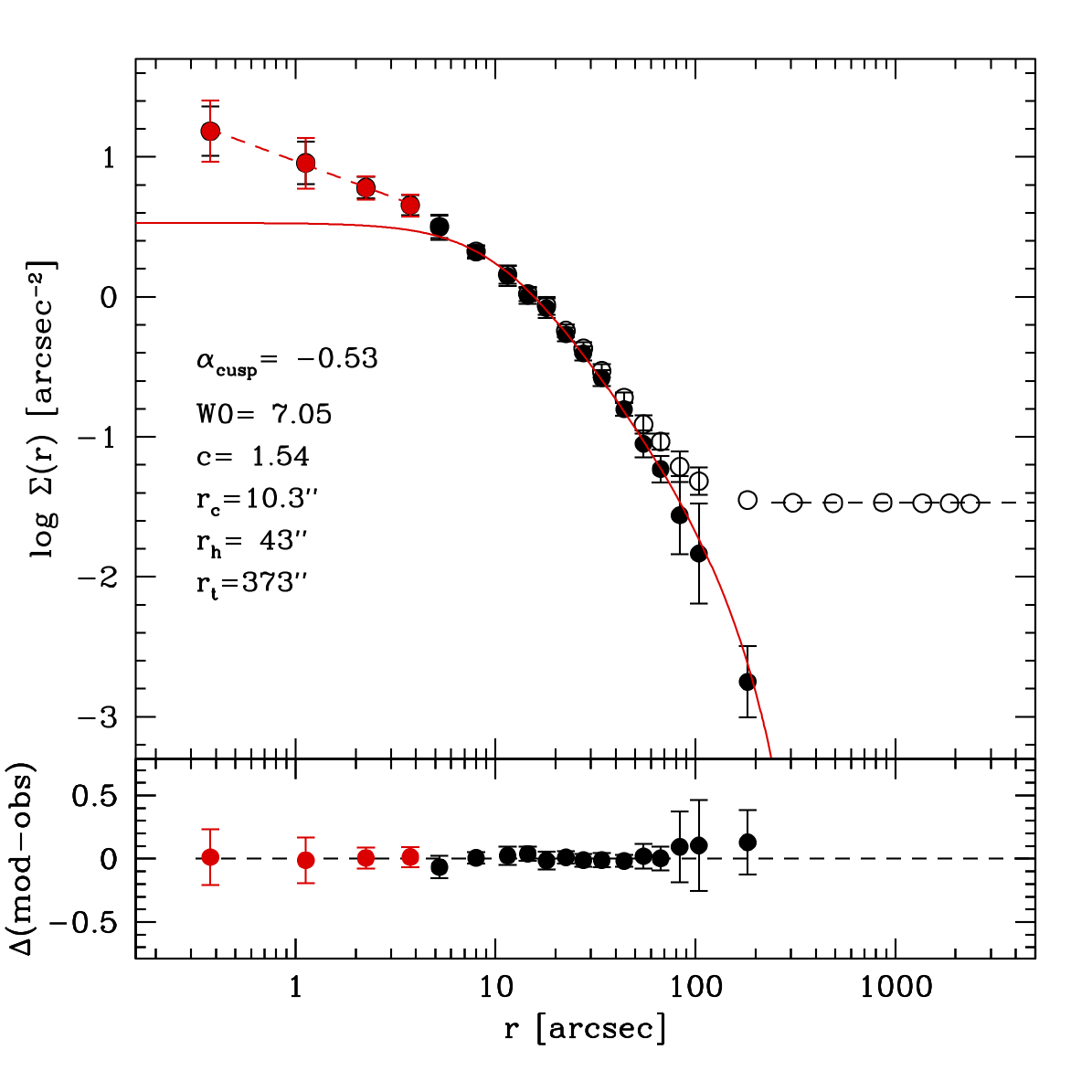}
\caption{Star density profile of NGC 6284. The left panel shows the observed profile (empty circles) and the background-subtracted profile (filled circles). The dashed horizontal line is the background density value used to decontaminate the cluster
  profile from the Galactic field contribution. The solid red curve is the King model that best fits the innermost portion of the profile, obtained for illustrative purposes by artificially increasing the errors of the outer points. The corresponding values of some structural parameters are labelled. The bottom panel shows the residuals between the best-fit King model and the cluster density profile. In the right panel, the circles are the same as in the left panel, but the four innermost points are coloured in red to highlight the presence of a central density cusp. These have been fitted with the power-law function shown as a red dashed line, with slope $\alpha_{\rm cusp}=-0.53$. The best-fit King model to the profile beyond the four inner points (black solid circles) is shown with the red solid curve, and is characterised by the labelled structural parameters.}
\label{fig:density_profile}
\end{center}
\end{figure*} 

\section{Observations and data reduction}
\label{sec:photometry}
The present study is based on a data set obtained with the Wide Field
Camera 3 (WFC3) onboard the $HST$ (GO: 15232, PI: Ferraro). The data
set comprises seven images ($6 \times27$ s and 1$\times$ 666 s
exposures) in the F555W filter, and seven images ($6\times15$ s and $1
\times $ 617 s exposures) in the F814W filter.

The photometric analysis has been carried out on the -flc images,
which are already corrected for dark-subtraction, flat-field, bias,
and charge transfer efficiency. We have used the standard package
\textrm{DAOPHOT IV} \citep{Stetson1987} following the prescriptions by
\citet{Cadelano2019, Cadelano2020a}.  We start by modelling the shape
of the Point Spread Function (PSF) by selecting 200 isolated stars
present in each image. For the WFC3 camera, the Full Width at Half
Maximum (FWHM) is set at 1.5 pixels ($\sim0\arcsec.06$), and we
sampled each of the selected stars within a 20-pixel radius
($\sim0\arcsec.8$). For each image, the best PSF model was provided by
a Moffat function \citep{Moffat1969} and a Penny function
\citep{Penny1976} for the F555W and the F814W filters, respectively,
based on a $\chi^2$ statistic. These models were applied to all the
sources detected in our images above a 5$\sigma$ threshold from the
local background level. We then built a master catalogue containing the
instrumental magnitudes and positions of each stellar source detected
in at least 3 images. At the corresponding positions of all these
sources, a fit was forced in each image using
\textrm{DAOPHOT/ALLFRAME} \citep{Stetson1994}. The results have been
combined together using \textrm{DAOPHOT/DAOMASTER} to finally obtain
homogenised instrumental magnitudes and their related photometric
errors.  The instrumental magnitudes have been calibrated to the
VEGAMAG system using the latest zero points derived by
\citet{Calamida2022} and reported at the $HST$ WFC3
website\footnote{\url{https://www.stsci.edu/hst/instrumentation/wfc3/data-analysis/photometric-calibration/uvis-photometric-calibration}},
namely, ZP$_{\rm F555W1}$ = 25.838 and ZP$_{\rm F555W2}$ = 25.825 in
the F555W filter for the stars detected in chips 1 and 2,
respectively, and ZP$_{\rm F814W1}$ = 24.699 and ZP$_{\rm F814W2}$ =
24.684 for those detected in chips 1 and 2 in the F814W
filter. Finally, we applied independent aperture and encircled energy
corrections for each chip and filter.  The instrumental positions of
the stars present in our images were corrected for geometric
distortions in both chips, following the procedure described in
\citet{Bellini2011}. We then converted these updated instrumental
positions into the absolute system celestial coordinates (RA and Dec)
through cross-correlation with the stars in common with the $Gaia$-DR3
catalogue \citep{Gaia2022}.

The star catalogue thus obtained allowed us to generate the first deep
CMD of NGC 6284 extending more than 6 magnitudes below the Main
Sequence Turn Off (MS-TO; see Fig. \ref{fig:cmd_complete}). For
comparison, the previously available CMDs reached just $\sim$2 mag
below the horizontal branch in the $J$ and $K$ filters
\citep{Minniti1995}, and $\sim$2 mag below the MS-TO in the F439W and
F555W filters \citep{Piotto2002}.  As mentioned in Section
\ref{sec:intro}, the sky position of the cluster toward the outskirts
of the bulge causes a mild level of contamination of the CMD from
field interlopers. Unfortunately, at the moment, a cluster membership
selection cannot be applied with the available data sets, neither
through proper motions, nor through a statistical decontamination (see
e.g. \citealt{Dalessandro2019, Cadelano2017a}).  Nevertheless, the main features of the CMD are
easily distinguishable. The MS spans a range of magnitudes between
$21.0 < m_{\rm F555W}< 26.5$ and shows the MS-TO point at around
$m_{\rm F555W} = 20.8$. The CMD also shows a horizontal branch at
about $m_{\rm F555W} = 17.3$ that is mainly populated in the blue
side, well-defined sub-giant and red giant branches (SGB and RGB,
respectively) with the characteristic RGB-bump
\citep{fusi+90,ferraro+99a} at $m_{\rm F555W} = 17.5$, and a rather
tenuous asymptotic giant branch above the horizontal branch.

\begin{figure*}[ht!]
\includegraphics[scale=0.41]{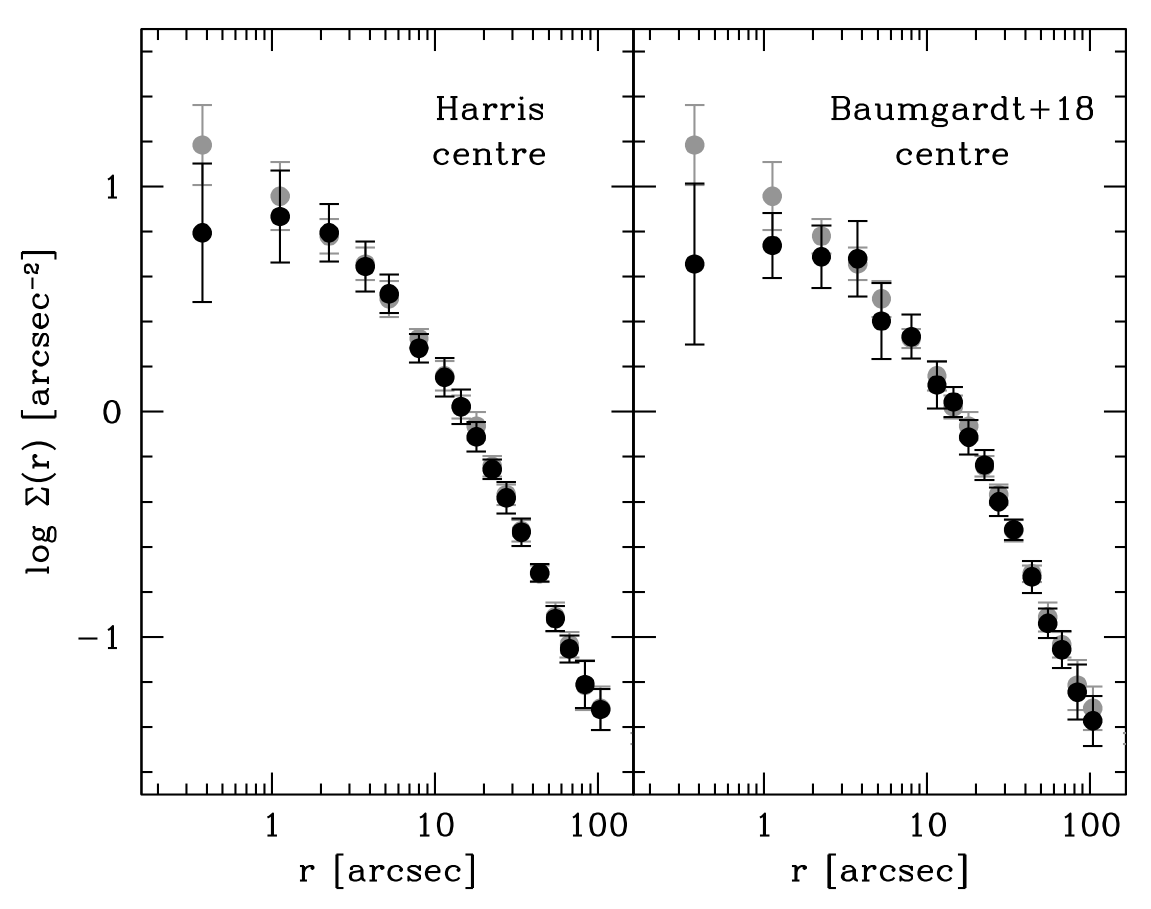} 
\caption{The density profile obtained in this
  work, calculated with respect to the value of $C_{\rm grav}$ quoted
  in Section \ref{sec:cgrav}, is shown with grey circles (same as the
  empty circles in Fig. \ref{fig:density_profile}). Left and right panels show the projected density profile  calculated with respect to the
  cluster centres quoted in the \citet{Harris1996} catalogue and in \citet{Baumgardt2018}, respectively (black circles). Only the inner
  portion of the profile, out to $r\sim 100\arcsec$, is shown in the
  figure.
\label{fig:cfr_dens}}
\end{figure*}
 
\section{The gravitational centre and star density profile of NGC 6284}
\label{sec:rad_prof}

\subsection{Centre of gravity}
\label{sec:cgrav}
To build the projected density profile and to estimate the structural
parameters of NGC 6284, we first need an accurate determination of its
gravitational centre ($C_{\rm grav}$).  The presence of a few luminous
stars in the surveyed field can generate an offset between the surface
brightness peak and the true gravitational centre of the cluster
(``shot-noise bias''). Hence, we estimated $C_{\rm grav}$ from the
position of resolved stars, following the procedure already employed
in many previous works \citep[e.g.,][]{Montegriffo1995, Lanzoni2007b,
  Lanzoni2019, Raso2020}. Briefly, starting from a first-guess centre
(the one reported in the catalogue of Orbital Parameters of Galactic
Globular Cluster; \citealp{Baumgardt2018}), we iteratively calculated
the average of the projected $x$- and $y$-coordinates of a sample of
stars selected above a given limiting magnitude (bright enough to
avoid incompleteness issues) and within a cluster-centric distance
larger than the literature core value (to sample the radial region
where the density profile starts to decrease). The procedure stops
when 10 consecutive iterations yield values that differ by less than
0.01$\arcsec$.

We repeated the computation for three different magnitude limits
($m_{\rm F555W} = 20.4, 20.7$ and 21.2) and for cluster-centric
distances ranging from $25\arcsec$ to $36\arcsec$ in steps of
$2\arcsec$, $3\arcsec$, and $5\arcsec$.  The final coordinates of
$C_{\rm grav}$ correspond to the average of the centres determined for
each combination of selection magnitudes and radii, and they turn out
to be R.A. $= 256.1192583^\circ$ ($17^{\rm h}$ $04^{\rm m}$
$28.622^{\rm s}$) and Dec = $-24.7648085^\circ$ ($-24^{\circ}$
45$\arcmin$ 53.31$\arcsec$), with an uncertainty of $\sim
0.1\arcsec$. Our estimate is located $\sim 2.9\arcsec$ west from the
first-guess centre \citep{Baumgardt2018}, and $\sim 1.5\arcsec$ east
from that quoted in the \citet{Harris1996} catalogue.

\subsection{Stellar density profile}
The shot-noise bias mentioned above may also have a significant impact
on the shape of the surface brightness profile and, as a consequence,
on the cluster structural parameters estimated from it \citep[see,
  e.g.][]{Noyola2006}. To circumvent this issue, we have promoted
\citep{Montegriffo1995, ferraro+99b, Ferraro2003} and then
systematically adopted the use of resolved star counts to construct
projected star density profiles and to determine the cluster structural
parameters \citep[e.g.,][]{Lanzoni2007a, Lanzoni2007b, Lanzoni2007c,
  ibata+09, Miocchi2013}.

To construct the stellar density profile of NGC 6284, we combined our
WFC3 photometric catalogue (which extends out to $\sim 115\arcsec$ from
the newly determined centre) with a $Gaia$-DR3 \citep{Gaia2022}
catalogue covering a circular area with a radius of 1$^\circ$. This
allowed us to accurately sample both the innermost and the outermost
regions of the cluster. We followed the procedure described in
\citet[][see also, e.g., \citealp{Cadelano2017b, Lanzoni2019,
    Raso2020}]{Miocchi2013}, selecting stars brighter than the MS-TO
in both catalogues: to this end, a magnitude cut at $m_{\rm F555W}<21.0$
has been applied to the WFC3 catalogue, while $G_{\rm mag}<20.0$ has
been adopted for the $Gaia$ one. Such a selection is made to include
only stars with approximately the same mass and to avoid spurious
effects from photometric incompleteness. For the WFC3 catalogue we then
considered 17 concentric rings around $C_{\rm grav}$ with radii up to
$115\arcsec$. This allowed us to sample the inner portion of the
density profile. In the case of the $Gaia$-DR3 data, we used 8
concentric rings (always centred around $C_{\rm grav}$) with radii
between $38\arcsec$ and $3600\arcsec$ to sample the outer cluster
regions, as well as the Galactic background density.  We then divided
each ring into three subsectors and determined the surface density in
each of them as the ratio between the number of stars and the
subsector area.  Finally, we adopted the average and standard
deviation of the density measurements in each subsector as the
resulting density value of the ring and its related uncertainty,
respectively.  The $Gaia$ profile was then vertically rescaled to
match that of the WFC3 using 5 points in the common radial range
(between $38\arcsec$ and $115\arcsec$). This provided us with the full
stellar density profile of the cluster that is plotted with empty
circles in Fig. \ref{fig:density_profile}.

\begin{figure}[ht!]
\includegraphics[scale=0.41]{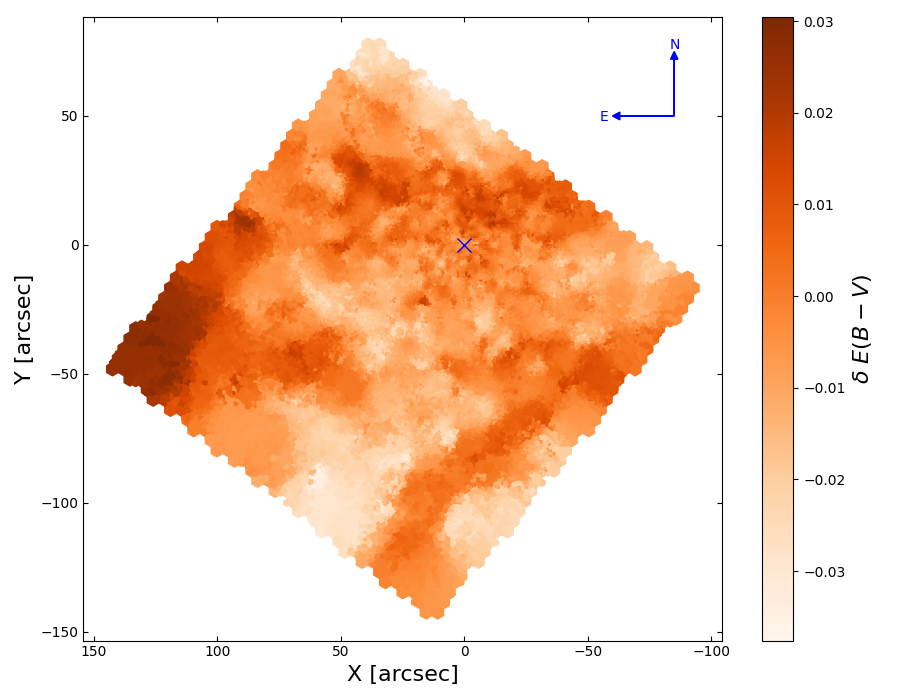}
\caption{Differential reddening map of NGC 6284 within the field of
  view sampled by the WFC3 images. Lighter colours correspond to less
  extincted areas (see the colour bar). The blue cross marks the
  cluster $C_{\rm grav}$ and the coordinates along the x- and y-axes
  are referred to it.}
\label{fig:redmap}
\end{figure}

Clearly, the stellar density is not constant in the innermost region,
at odds with the expectation from the \citet{King1966} model behaviour
observed in most GCs.  Indeed, the stellar density progressively
increases toward the centre of NGC 6284, drawing a steep cusp.
As discussed in the Introduction, this is the typical signature
of a GC that has undergone core collapse.
The projected density keeps decreasing for increasing distance from
the centre out to $r\sim 200\arcsec$, where the Galactic field becomes
dominant over the cluster population.  On the surveyed spatial scale
the distribution of background stars is expected to be approximately
uniform, thus explaining the well-defined density plateau visible in
the outermost portion of the profile (six outer empty circles in
Fig. \ref{fig:density_profile}).  To obtain the intrinsic density
profile of the system, the contribution from field stars must be
quantified and removed.  To this end, the level of Galactic field
contamination has been estimated by averaging the six values on the
plateau (see the horizontal dashed line in
Fig. \ref{fig:density_profile}), and then subtracted from the observed
distribution (empty circles). The field-subtracted star density
profile of NGC 6284 thus obtained is marked by the filled circles in
the figure.  It remains virtually unchanged at small radii because at
these distances from the centre the cluster population is by far
dominant. Conversely, the subtracted profile progressively diverges
from the observed one as the radius increases because the cluster
population decreases in number, while the field contribution remains
constant and starts to dominate.  This highlights the importance of
accurately quantifying the background level for a reliable
determination of the true density profile of stellar systems.

Given that the profile has been determined by using stars of
approximately the same mass, we can now fit it with single-mass
\citet{King1966} models to derive the structural parameters of the
cluster under the assumption of spherical symmetry and orbital
isotropy, as it is commonly done for GCs.  Of course, in the presence
of a central density cusp, a proper fit to the entire profile with
core-like models cannot be obtained \citep[e.g.,][]{Ferraro2003,
  Zocchi2016}.  Nevertheless, it is customary
\citep[e.g.,][]{Harris1996, Meylan1997} to fit the density (or surface
brightness) distribution of PCC clusters with King models of high
concentration parameter ($c\ge 2-2.5$) and small core radius
($r_c$). For illustrative purposes, we therefore forced the King model
fitting by artificially increasing the errors of the most external
points, thus giving larger weight to the innermost values. As
expected, the best-fit is provided by a model with $c=2.55$ and
$r_c=2.8\arcsec$ (see the left-hand panel of
Fig. \ref{fig:density_profile}). While it reasonably reproduces (by
construction) the innermost portion of the profile, it clearly fails
in properly fitting also the external part.

The only way to obtain an acceptable match of the entire star density
profile is by combining a King model in the outer regions, with a
power-law function in the centre. Indeed the innermost region ($r <
4.5\arcsec$, corresponding to the first 4 points of the profile) can
be nicely modelled by a straight line with slope $\alpha_{\rm cusp} =
-0.53 \pm 0.2$.
Then after excluding this portion, the profile is well reproduced by a
King model with $c=1.54$ and $r_c=10.3\arcsec$. The best-fit solution
has been searched by using a MCMC approach implemented by the \texttt{emcee} package \citep{Foreman2013, Foreman2019}, following
\citet{Cadelano2022}.  We assumed uniform priors on the parameters of
the fit (i.e., the King concentration parameter $c$, the core radius
$r_c$, and the value of the central density). Therefore, the posterior
probability distribution functions are proportional to the likelihood
$L = \exp(-\chi^2/2)$, where the $\chi^2$ statistic is calculated
between the measured density values and those predicted by the whole
family of adopted models.

\begin{figure*}[ht!]
\begin{center}
\includegraphics[scale=0.4]{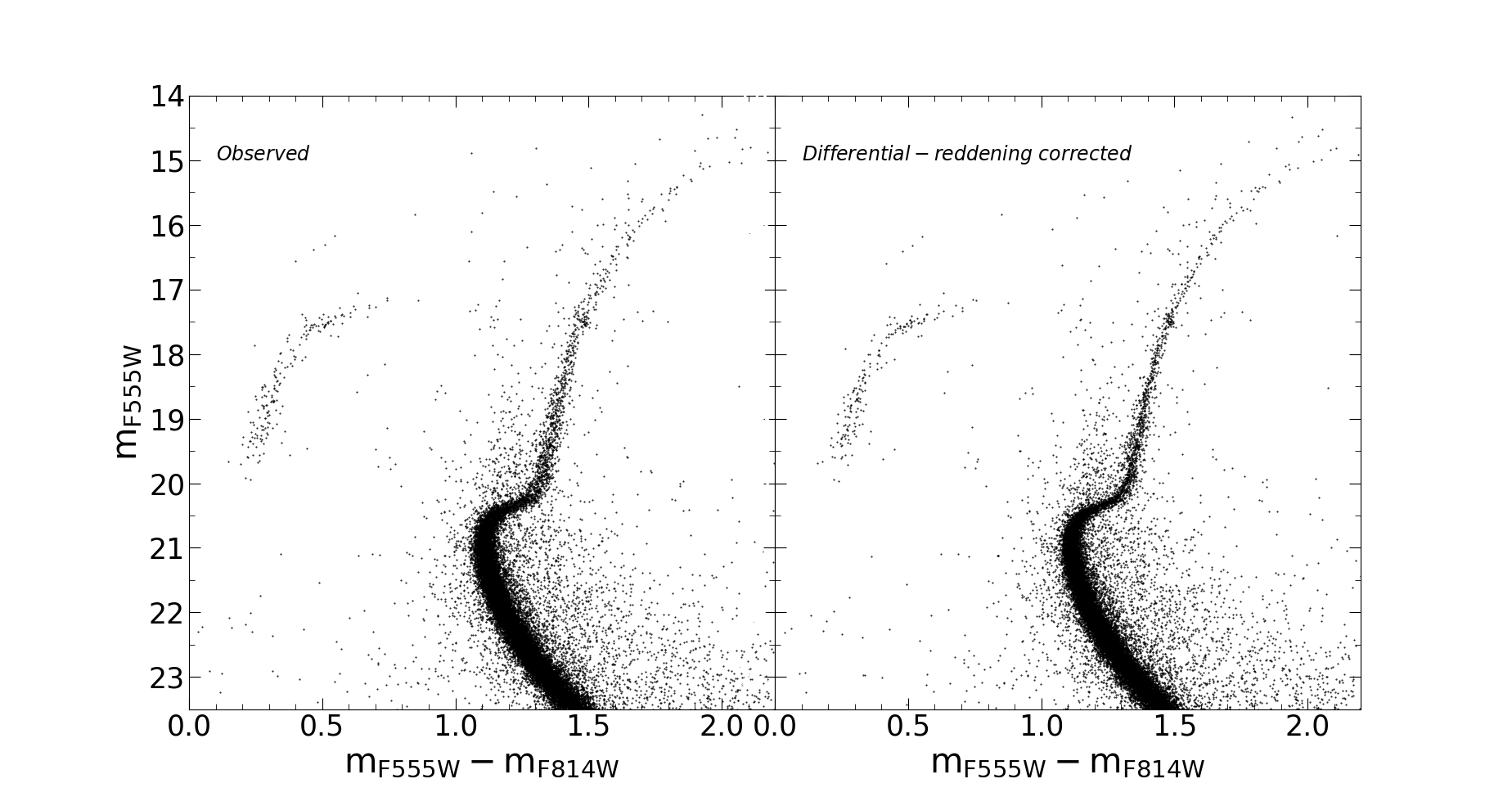}
\caption{Comparison between the original CMD obtained from the $HST$ WFC3 observations used in this work (left panel), and the CMD after it has been corrected for differential reddening (right panel).}
\label{fig:cmd_debv}
\end{center}
\end{figure*}

As a final note, Figure \ref{fig:cfr_dens} shows the comparison
between the observed density profile obtained in this work (grey
circles), and those derived with respect to the centres quoted in the
\citet{Harris1996} catalogue and in \citet{Baumgardt2018}, which are
offset, respectively, by $\sim 1.5-3\arcsec$ from our estimate of
$C_{\rm grav}$.  This nicely illustrates the importance of correctly
determining the cluster centre, especially in the case of very
concentrated GCs: indeed, a mismatch of just a few arcseconds makes
the central density cusp disappear.
 
\section{Differential reddening correction}
\label{sec:diff_red}
Although the \citet{Harris1996} catalogue quotes a moderate colour excess
for NGC 6284, $E(B-V) = 0.28$, and the evolutionary sequences in the
CMDs of Fig. \ref{fig:cmd_complete} appear only mildly distorted or
broadened, the fact that the cluster lies in the direction of the
bulge outskirts naturally recommends to quantify the possible effects
of differential reddening. We therefore determined a detailed
reddening map toward the system, making use of the iterative method
fully described in \citet{Pallanca2019, Pallanca2021a} and
\citet{Cadelano2020b}.

Briefly, we start by generating the CMD of a reference sample
including likely member stars, i.e., stars located at $r<30\arcsec$
from the centre of the cluster (to minimise the possible contamination
from field interlopers) and in a fixed magnitude range ($15.0 < m_{\rm
  F555W} < 23.0$) to include well-measured stars along the MS, SGB,
and RGB. We then divided this CMD into magnitude bins of 0.3 mag each,
except at the level of the MS-TO and SGB ($20.0 < m_{\rm F555W} <
21.25$) where we used 0.05 mag bins to obtain a finer sampling. For
each bin, we estimated the 3$\sigma$-clipped median values of the
($m_{\rm F555W} - m_{\rm F814W}$) colour and of the $m_{\rm F555W}$
magnitude. These medians were then interpolated to determine the mean
ridge line, which was used to estimate the geometric distance $\Delta
X$ of all the stellar sources in the reference sample, along the
direction of the reddening vector. By assuming the standard extinction
coefficient $R_V = 3.1$, this vector is defined by the extinction
coefficients $R_{\rm F555W} = 3.227$ and $R_{\rm F814W} = 1.856$,
obtained from \citet{Cardelli1989} and \citet{Girardi2002}.  For every
source in our catalogue, we then identified the $n$ closest reference
stars and computed the $\sigma$-clipped median of their geometric
distances: the result is the $\Delta X$ value finally assigned to the
source, which is transformed into the relative differential reddening
$\delta E(B-V)$ by using eq.(1) in \citet{Deras2023}.
To increase the spatial resolution we iteratively performed this
computation three times using the $n=100$, 50, and 25 closest stars.

This allowed us to generate the reddening map shown in Figure
\ref{fig:redmap}, which clearly shows the inhomogeneity of the medium
causing differential reddening across the cluster. The differential
variations of the colour excess are admittedly modest, ranging between
$-0.04 < \delta E(B-V) < 0.03$ within the sampled field of
view. Nevertheless, the broadening of the main evolutionary sequences
is significantly reduced and the main features in the CMD look sharper
after correction for differential reddening (see
Fig. \ref{fig:cmd_debv}). We will use this corrected CMD in the
subsequent analyses of this work.

\begin{figure*}[]
\begin{center}
\includegraphics[scale=0.4]{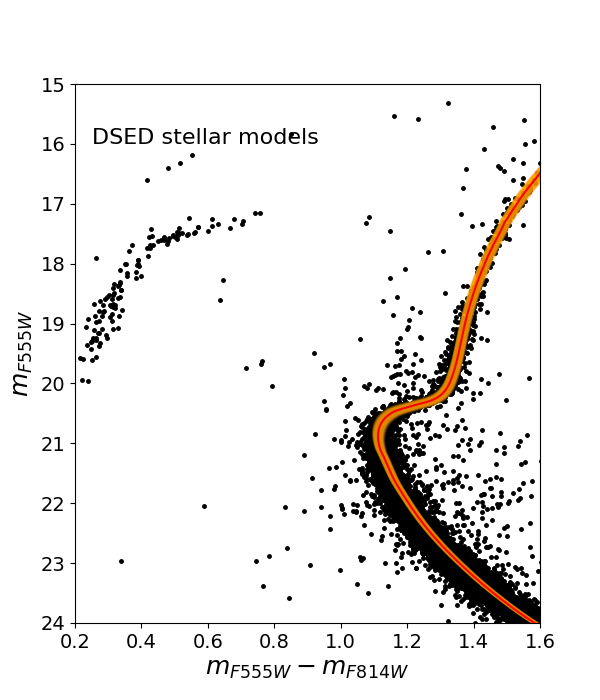}
\includegraphics[scale=0.25]{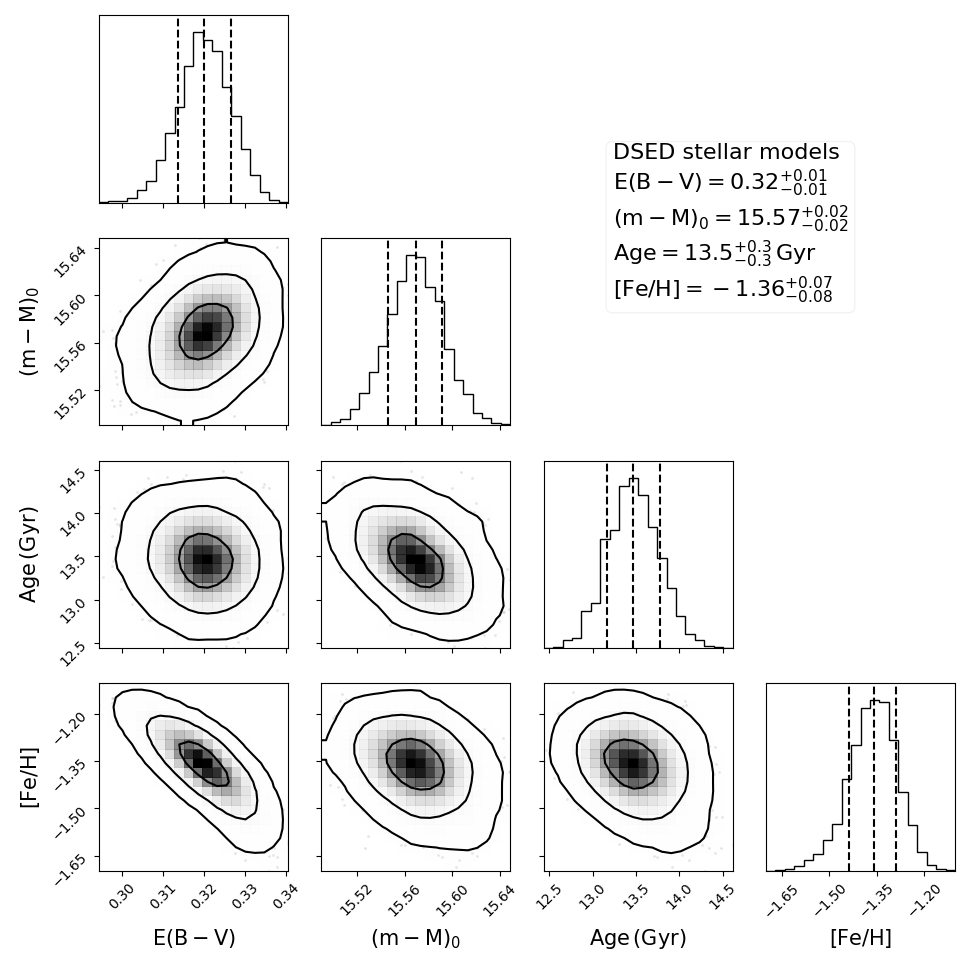}

\includegraphics[scale=0.4]{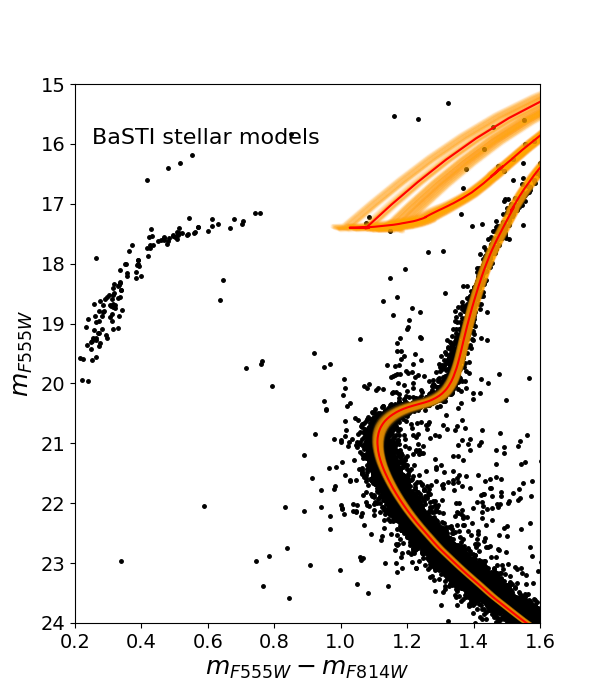}
\includegraphics[scale=0.25]{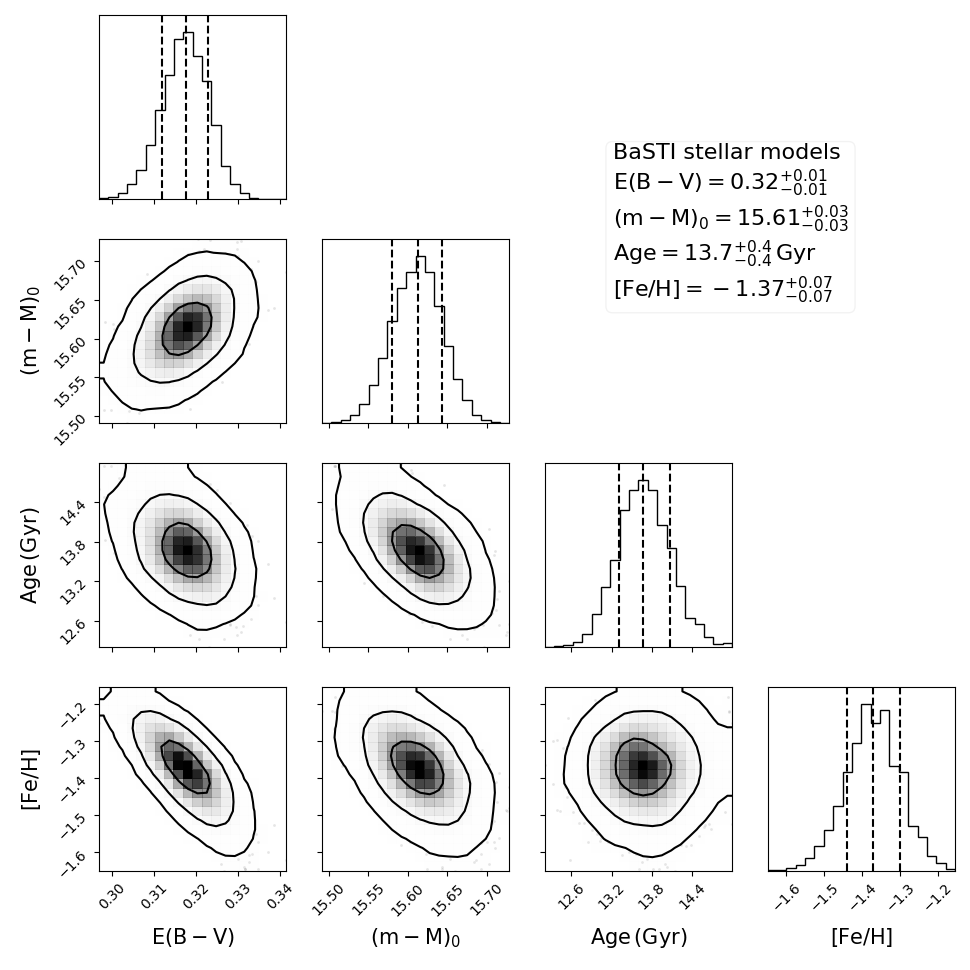}

\includegraphics[scale=0.4]{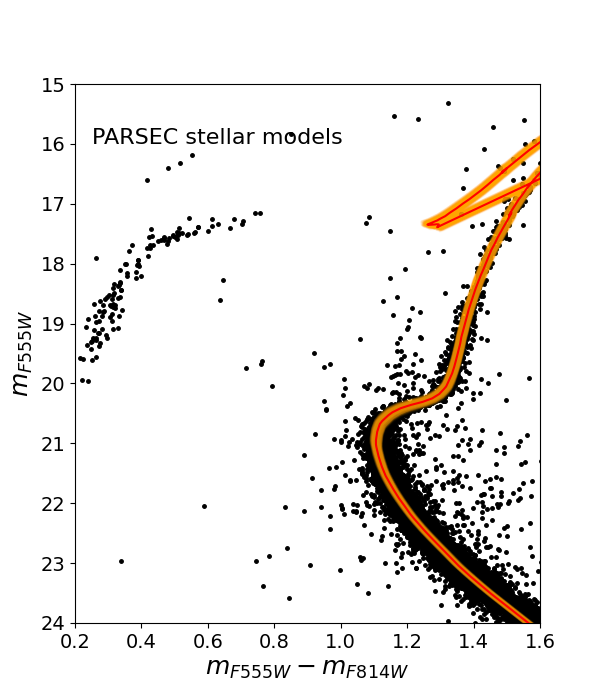}
\includegraphics[scale=0.25]{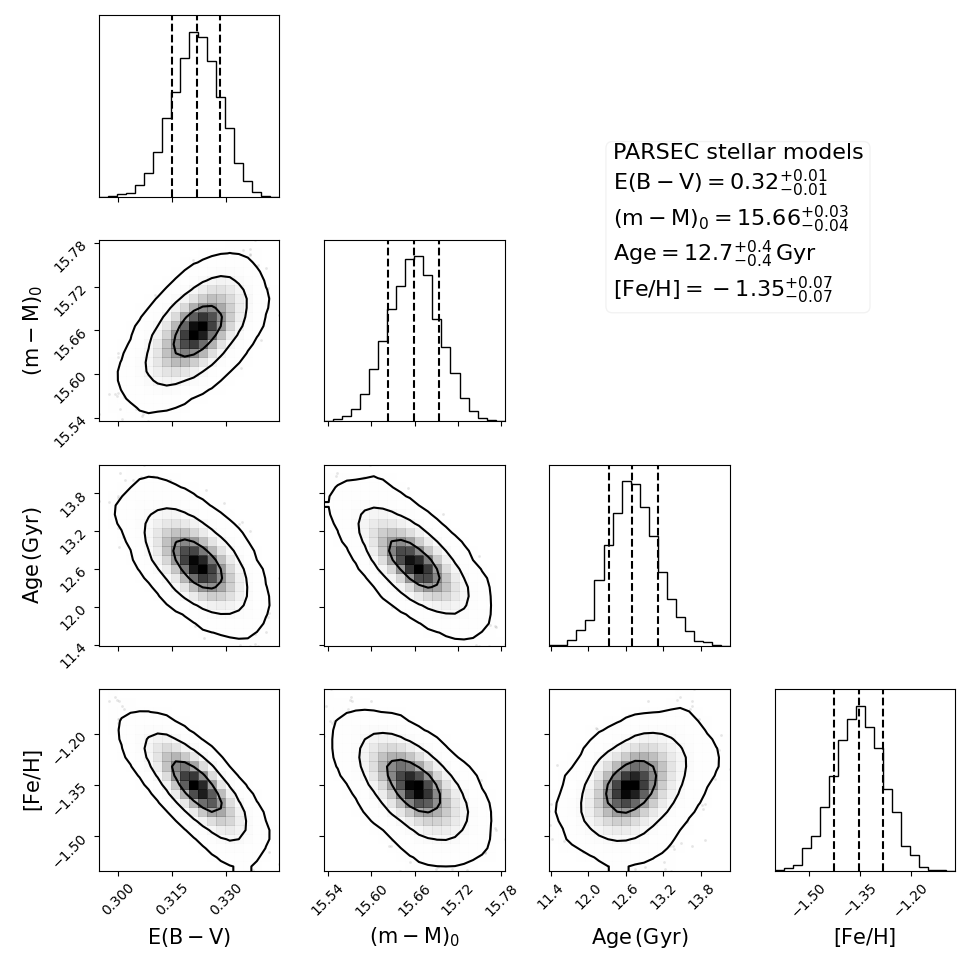}

\caption{
Isochrone fitting of the CMD of NGC 6284 corrected for differential reddening. Left panels show the best-fit corresponding to the DSED,
BaSTI and PARSEC isochrones (top, middle and bottom panel, respectively) plotted as a red solid line, and the $1\sigma$ envelope shaded in orange.
Right panels show the corner plots with the one- and two-dimensional projections of the posterior probability distributions for all the parameters
as obtained from DSED, BaSTI and PARSEC isochrones (top, middle and bottom panel, respectively). The contours correspond to the $1\sigma$, $2\sigma$, and $3\sigma$ levels}
\label{fig:isochrones}
\end{center}
\end{figure*}

\begin{table*}[ht!]
	\begin{center}
		\caption{Best-fit values of age, metallicity, colour
                  excess and absolute distance modulus obtained from
                  isochrone fitting.}\label{tab:iso_best}
		\begin{tabular}{|c|c|c|c|c|} 
			\hline
			Model	&   Age    & $[\rm Fe/H]$ & $E(B-V)$   & $(m-M)_{0}$ \\
				& [Gyr]   &   dex &    [mag]    &  [mag]     \\			
			\hline
			DSED   &  $13.5^{+0.3}_{-0.3}$ & $-1.36^{+0.07}_{-0.08}$ & $0.32^{+0.01}_{-0.01}$ & $15.57^{+0.02}_{-0.02}$  \\ 
			BaSTI  &  $13.7^{+0.4}_{-0.4}$ & $-1.37^{+0.07}_{-0.07}$ & $0.32^{+0.01}_{-0.01}$ & $15.61^{+0.03}_{-0.03}$  \\ 
			PARSEC &  $12.7^{+0.4}_{-0.4}$ & $-1.35^{+0.07}_{-0.07}$ & $0.32^{+0.01}_{-0.01}$ & $15.66^{+0.03}_{-0.04}$  \\ 
			\hline
			 Average $\pm~\sigma$  & $13.3 \pm 0.4$ & $-1.36 \pm 0.01$ & $0.32 \pm 0.01$ & $ 15.61 \pm 0.04$\\
			\hline
		\end{tabular}
	\end{center}
\end{table*} 
 
\section{Isochrone fitting}
\label{sec:age}
To simultaneously obtain a photometric estimate of the absolute age,
metallicity, distance modulus, and absolute average reddening of NGC
6284, we have applied a Bayesian procedure similar to that used by
\citet[][see also \citealp{Saracino2019, Cadelano2019, Cadelano2020a,
    Deras2023}]{Cadelano2020b}, which consists in performing an
isochrone fitting of the CMD corrected for differential reddening. The
isochrones were extracted from three different databases, namely, DSED
\citep{Dotter2008}, BaSTI \citep{Pietrinferni2021}, and PARSEC
\citep{Marigo2017}. For each isochrone, we assumed a standard He mass
fraction ($Y$ = 0.25), and [$\alpha$/Fe]$= +0.4$, which is a typical
value for Galactic GCs.  To minimise the contamination from field
interlopers and further improve the accuracy of our results, we only 
considered stars within $50\arcsec$ from $C_{\rm grav}$ (see
Section \ref{sec:rad_prof}), with high-quality photometry (i.e.,
sharpness parameter $|sh|\leq0.025$) and observed in the MS-TO, SGB,
and RGB up to the RGB-bump, i.e., in the magnitude range $17.5< m_{\rm
  F555W}<21.0$.  In fact, this is the region most sensitive to stellar
age and metallicity variations. A grid of isochrones was computed in a
wide range of ages (from 10.0 to 15.0 Gyr), metallicities (from
[Fe/H]$=-2.10$ to [Fe/H]$=-0.90$), distance moduli (between 15.00 and
17.00), and colour excess $E(B-V)$ (between 0.10 and 0.50).
The comparison between the CMD and the adopted grid of isochrones has
been performed by using the MCMC sampling technique, following the
same computational approach described in \citet[][see their Section
  4.2]{Cadelano2020b}. To sample the posterior probability
distribution in the $n$-dimensional parameter space, we used the
\texttt{emcee} code \citep{Foreman2013,Foreman2019}. For reddening and
distance modulus we adopted Gaussian prior distributions peaked at
$E(B-V) = 0.28 \pm 0.03$ \citep{Harris1996} and $(m-M)_0 = 15.76 \pm 0.06$ \citep{Baumgardt2021},
respectively. For the age and metallicity, we assumed flat priors in the adopted range of values. When converting absolute magnitudes to
the observed frame, we adopted the temperature-dependent extinction
coefficients from \citet{Casagrande2014}.

The left panels of Figure \ref{fig:isochrones} show the CMD and the
best-fit isochrones with their 1$\sigma$ uncertainty for the three
adopted sets of theoretical models (see labels).  The one- and
two-dimensional posterior probabilities for all of the parameter
combinations are presented on the right panels as corner plots.  Table
\ref{tab:iso_best} summarises the best-fit values of the four
parameters and their uncertainties (based on the $16^{th}$, $50^{th}$,
$84^{th}$ percentiles).  The results obtained from the different sets
of models are consistent within the uncertainties, and we therefore
adopted their average as final best-fit estimate of the four
parameters: namely, $t=13.3 \pm 0.4$ Gyr, [Fe/H]$ = -1.36 \pm 0.01$,
$E(B-V) = 0.32 \pm 0.01$, and $(m-M)_0 = 15.61 \pm 0.04$
(corresponding to a distance of 13.2 kpc).

\section{The RGB-bump}
\label{sec:bump}
One of the most notable features along the RGB is the so-called
RGB-bump. This evolutionary feature flags the moment when the
hydrogen-burning shell reaches the hydrogen discontinuity left by the
innermost penetration of the convective envelope (see the seminal
works by \citealt{fusi+90, ferraro+91, ferraro+92,
  ferraro+99a,ferraro+00}; see also the compilations by
\citealt{zoccali+99, valenti+04}, and more recently by
\citealt{nataf+13}).

\begin{figure} 
\begin{center}
\includegraphics[width=95mm]{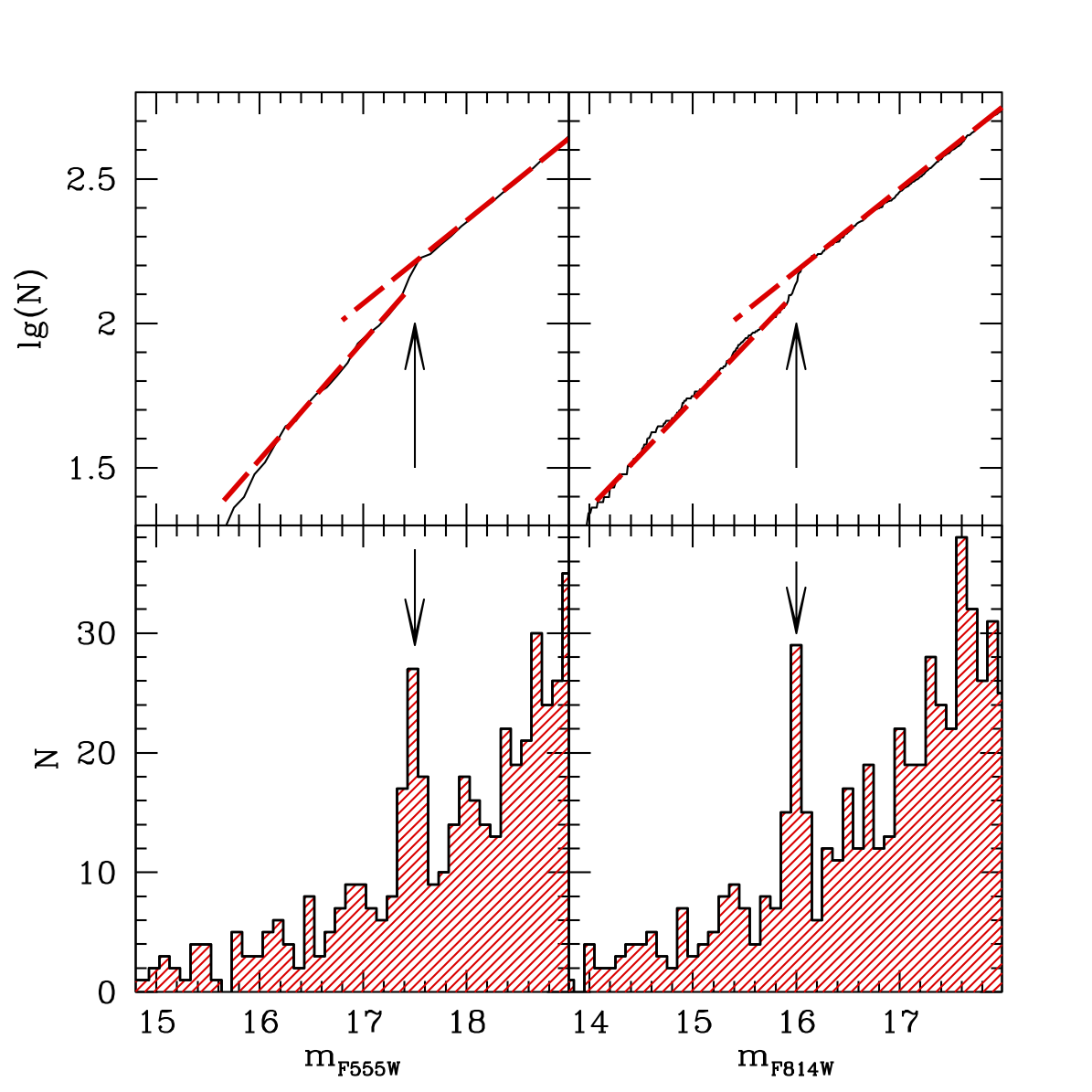}
\caption{Integrated (top panels) and differential (bottom panels)
  luminosity functions of RGB stars in the F555W and F814W filters
  (left and right panels, respectively). The two vertical arrows mark
  the change in the slope (top) and the local peak (bottom) unequivocally
  indicating the occurrence of the RGB-bump. }
\label{Fig:bump1}
\end{center}
\end{figure}

During the RGB evolution, the luminosity of low-mass stars is provided
by hydrogen burning reactions in a shell that progressively moves
outwards. At the same time, the convective envelope recedes toward the
stellar surface, leaving behind chemically mixed material. This
translates in a constant hydrogen abundance down to the innermost
radial distance reached by convection, which is flagged by a sharp
chemical discontinuity (see, e.g., Fig. 5.15 of
\citealp{salariscassisi}).  The RGB-bump corresponds to the moment
when the hydrogen-burning shell passes through this
discontinuity. Hence, the predicted luminosity of the RGB-bump depends
on the parameters and physical processes regulating the ability of
convection to penetrate deep into the stellar interior (for instance,
the parameters affecting the stellar opacity, as the heavy element and
the helium abundances): in the case of a deeper penetration, the
hydrogen burning shell crosses at earlier times the chemical
discontinuity left by the convective envelope, and the bump occurs at
fainter magnitudes along the RGB.  The observational determination of
the RGB-bump luminosity therefore provides crucial constraints to
stellar evolutionary models, especially if measured in large samples
of stellar systems with different properties (e.g., with different
metallicities).

From the observational point of view, this event produces two main
signatures that are intimately linked to the physical processes
occurring in the stellar interior: (1) a bump in the RGB differential
luminosity function (i.e., an increase of star counts at the RGB-bump
luminosity) due to a temporary hesitation in the star evolutionary
path while climbing the RGB, and (2) a change in the slope of the RGB
integrated luminosity function due to a change in the evolutionary
speed when the burning shell passes from a region of variable hydrogen
abundance (moderate climbing speed), to a region of constant hydrogen
abundance (fast climbing speed). As discussed in many papers
\citep[e.g.,][]{fusi+90, ferraro+99a, ferraro+00}, the latter
observable is a powerful diagnostic especially in situations of poor
statistics, since it is obtained from incremental star counts over a
range of several magnitudes: this makes it much less prone to
statistical fluctuations with respect to the case of a differential
luminosity function, where stars are counted in discrete bins of
magnitude.

To add further contribution to the observational estimate of the
RGB-bump luminosity in clusters with different properties, here we
determined these two diagnostics in the case of NGC 6284.  From a
visual inspection of Fig. \ref{fig:cmd_debv}, the location of the
RGB-bump is easily recognisable in the CMD as a grouping of stars
about halfway along the RGB. However, to properly determine its
magnitude level, we selected the RGB population from the differential
reddening-corrected CMD and we determined its integrated and
differential luminosity functions (top and bottom panels of Figure
\ref{Fig:bump1}, respectively). A clear change in the slope in the
integrated luminosity function (due to a variation in the climbing
speed of stars along the RGB), and a star count excess in the
differential luminosity function (testifying their temporary
evolutionary hesitation) are well apparent in the figure, and they
allow the precise determination of the RGB-bump magnitude (see the
arrows): $m_{\rm F555W}=17.46\pm0.05$ and  $m_{\rm F814W}=16.00\pm0.05$.

\begin{figure} 
\begin{center}
\includegraphics[width=90mm]{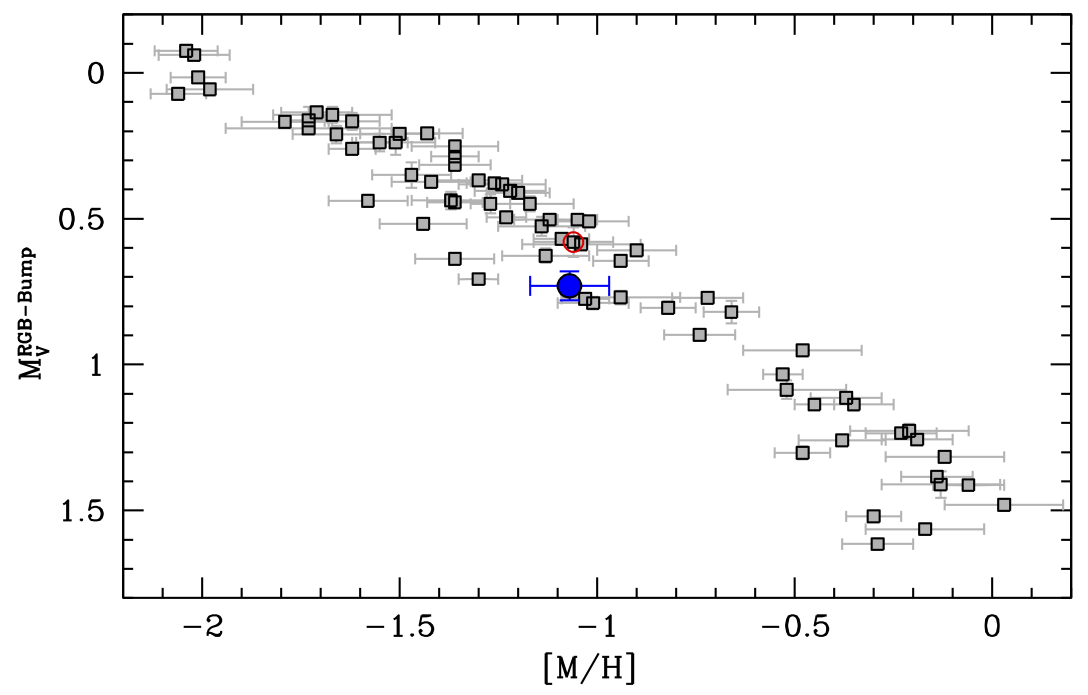}
\caption{Absolute magnitude of the RGB-bump in the Johnson $V$ band as
  a function of the global metallicity [M/H] for a large sample of
  Galactic GCs (grey symbols, from \citealp{nataf+13}).  The large
  blue circle marks the value determined in this work for NGC 6284,
  while the red empty circle marks the measure quoted in
  \citet{nataf+13}.}
\label{Fig:bump2}
\end{center}
\end{figure}

One of the parameters affecting the RGB-bump luminosity is the cluster
overall metallicity: for increasing metallicity, the stellar opacity
increases, the convective envelope penetrates deeper in the interior
of the star, and the RGB-bump therefore occurs at fainter magnitudes
along the RGB.
Indeed, a strong relation is observed between the metallicity and the
RGB-bump magnitude of Galactic GCs, a recent compilation (also
including NGC 6284) being presented in \citet{nataf+13}.  To compare
our determination with the values in this compilation, we first need
to convert our WFC3 F555W and F814W magnitudes into the standard
(Johnson) $V$ and $I$ magnitudes. According to \citet{harris+18}, the
following simple transformations can be adopted: $(V-m_{\rm
  F555W})_0=-0.093\times(V-I)_0$ and $(I-m_{\rm F814W})_0=0$, where
the subscript refers to reddening-corrected magnitudes.  We
therefore used the average colour excess estimated in the previous
section, $E(B-V) = 0.32$, to determine the reddening-corrected WFC3
magnitudes of the RGB-bump.  Then, by applying the transformations
above, we found $V_0^{\rm bump}=16.34$.  By adopting the true distance
modulus previously estimated, $(m-M)_0 = 15.61$, this corresponds to a
Johnson $V$-band absolute magnitude $M_V^{\rm bump}=0.73$, with an
uncertainty of about 0.05 magnitudes.  To determine the cluster global
metallicity we used the relation reported in \citet{salaris+93},
adopting [$\alpha$/Fe]$=+0.4$ and the photometric estimate of the iron
abundance derived in the previous section ([Fe/H]$= -1.36$), thus
finding [M/H]$=-1.07$.  Figure \ref{Fig:bump2} shows our estimate
(large blue circle) compared to the values listed in the
\citet{nataf+13} compilation. While fitting into the overall trend,
the value found in the present study is $\sim 0.2$ mag fainter than
that quoted by \citet[][which is highlighted by a red empty circle in
  the figure]{nataf+13}.

\section{Summary and conclusions}
\label{sec:summary}
By using high-resolution $HST$ observations, we obtained the first
high-quality CMD of NGC 6284, extending down to $\sim 6$ magnitudes
below its MS-TO. This allowed us to properly characterise the
innermost regions of this high-density and poorly studied GC, which is
located in the direction of the Galactic bulge outskirts.

For the first time, we determined the centre of gravity and the
density profile of the cluster from resolved star counts.
We found that $C_{\rm grav}$ is displaced by $1.5-3\arcsec$ from the
values reported in the literature \citep{Harris1996,
  Baumgardt2018}. The correct determination of the cluster centre
allowed us to clearly detect the presence of a steep power-law cusp in
the innermost portion ($r<4.5\arcsec$) of the projected density
profile, in agreement with previous findings from surface brightness
investigations that classified NGC 6284 as a PCC cluster
\citep[see][and references therein]{lugger+95}.
By complementing our $HST$ photometry with a wide-field catalogue from
$Gaia$-DR3, we determined the cluster density profile over its entire
radial extension and properly characterised (and then subtracted) the
contribution from the Galactic field, which starts to become
dominating at $r\sim 200\arcsec$ from the centre.  As expected, the
overall profile cannot be properly described by a King model, due to
the presence of the central cusp (see
Fig. \ref{fig:density_profile}). By forcing the fit in the innermost
portion of the profile, the best solution is a King model with very
high concentration parameter ($c=2.55$) and very small core radius
($r_c=2.8\arcsec$, corresponding to 0.18 pc at the distance determined
in this work). However, this model clearly overestimates the stellar
density in the cluster outskirts, bringing to unreliably large values of
the half-mass and tidal radii.  A proper fit to the entire profile
requires a two component model, consisting in a power-law with slope
$\alpha_{\rm cusp} = -0.53$ in the innermost region ($r<5\arcsec$),
and a single-mass King model with concentration $c=1.54$, core radius
$r_c=10.3\arcsec$ (0.66 pc) and half-mass radius $r_h=43\arcsec$ (2.75
pc) at larger distances. These are the typical features observed in
the case of highly evolved GCs, that already experienced core
collapse.

\begin{figure}[ht!]
\begin{center}
\includegraphics[width=90mm]{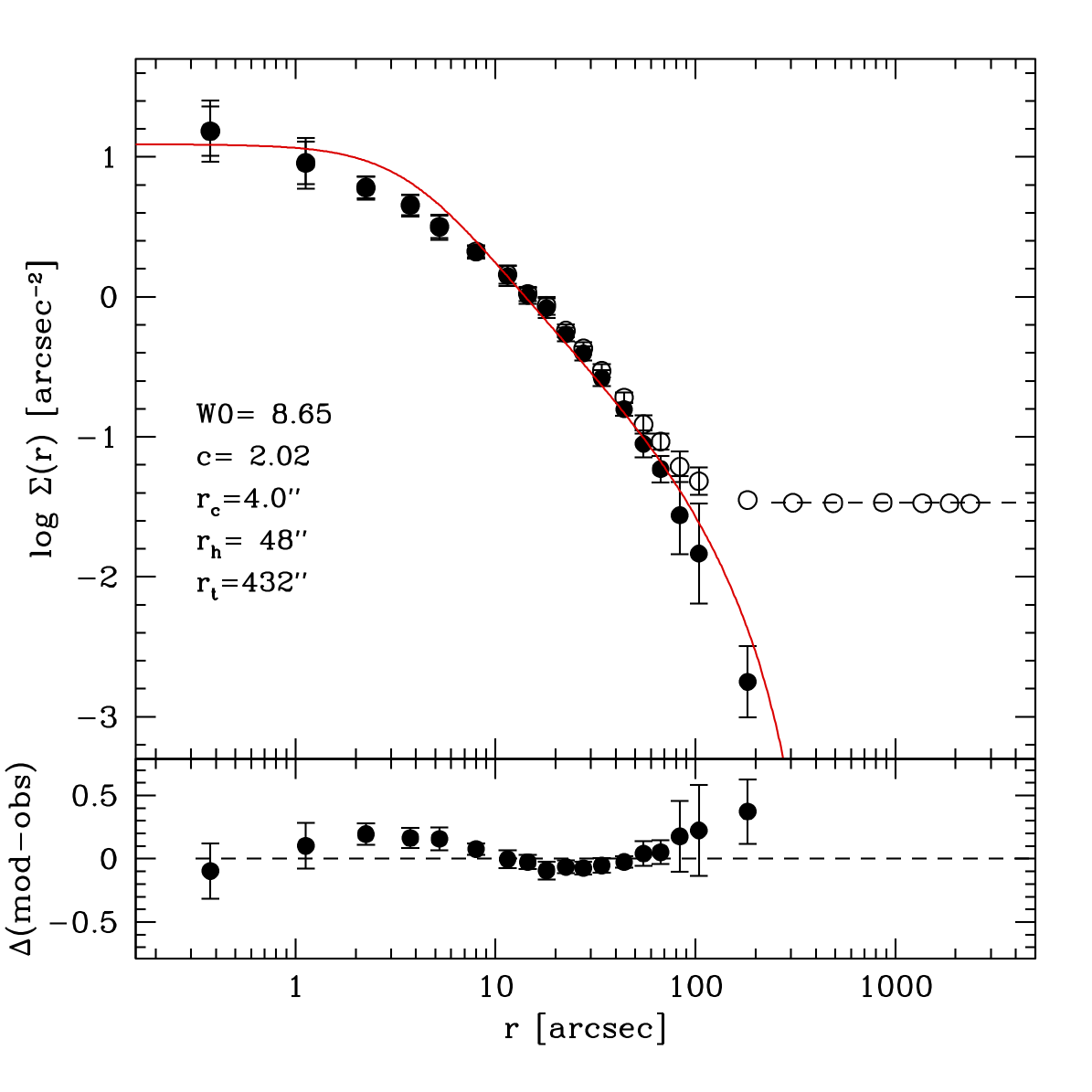}
\caption{As in Fig. \ref{fig:density_profile}, but with a King model
  (red line) that is aimed at approximately reproducing simultaneously
  both the inner and the outer regions of the profile.}
\label{fig:king_global}
\end{center}
\end{figure} 

The acquired observations also allowed us to determine a
high-resolution reddening map in the cluster direction, which shows
moderate inhomogeneities (Fig. \ref{fig:redmap}). In particular, it is
worth noticing the patchy semi-horizontal strip in the northern
hemisphere, near $C_{\rm grav}$.  We extracted isochrones from three
different sets of models (DSED, BaSTI, and PARSEC) and used them to
perform a fit to the MS-TO and RGB regions of the differential
reddening corrected CMD. With the caveat that a fit with four free
parameters may be prone to degeneracies, this allowed us to obtain
updated estimates of the absolute age, metallicity, distance modulus,
and reddening of this poorly studied cluster (see Table
\ref{tab:iso_best}).  Our estimate of the age yields $13.3 \pm 0.4$
Gyr, which is about 2 Gyr older than some values reported in the
literature (see Section \ref{sec:intro}), but consistent with the age
of most GCs in our Galaxy.  The derived a metallicity ([Fe/H]=$-1.36
\pm 0.01$) is consistent with the value measured back in 1984 from
low-resolution spectroscopy \citep[][see also Harris1996]{zinnwest84}.
The distance modulus $(m-M)_0 = 15.61 \pm 0.04$ translates into a
distance of $13.2 \pm 0.2$ kpc. This is $\sim 1$ kpc smaller than the
one reported by \citet[][14.2 kpc]{Baumgardt2021}, and $\sim 2$ kpc
smaller than the one quoted by \citet[][15.2 kpc]{Harris1996}, but
consistent with that obtained from RR Lyrae by \citet[][12.9
  kpc]{Clement1980}.  We estimate an absolute reddening $E(B-V) = 0.32
\pm 0.01$, which is slightly larger than the average value found in
the literature (0.28), but still consistent with it within the
typically adopted 10$\%$ uncertainty.

\begin{table}[ht]
\caption{General Parameters of NGC 6284.}
\begin{center}
\begin{tabular}{ l l}
\hline
\hline
Parameter & Estimated value \\
\hline
Centre of gravity               & $\alpha_{\rm J2000}= 17^{\rm h} 04^{\rm m} 28.6^{\rm s}$ \\
                                & $\delta_{\rm J2000} = -24^{\circ} 45\arcmin 53.31\arcsec$\\
Age                             & $t=13.3 \pm 0.4$ Gyr\\
Metallicity                     & [Fe/H] = $-1.36 \pm 0.01$ \\
Distance Modulus                & $(m-M)_{0} = 15.61 \pm 0.04$ \\
Reddening                       & $E(B-V) = 0.32 \pm 0.01$ \\
Central relaxation time         & $\log(t_{rc}/{\rm yr}) = 6.72$\\
Half-mass relaxation time       & $\log(t_{rh}/{\rm yr}) = 9.06$\\
\hline
\end{tabular}
\end{center}
\label{table:tab_params}
\end{table}

The high quality of the CMD derived in this work also allowed the
clear identification of the RGB-bump. Based on the integrated and
differential RGB luminosity functions (Fig.\ref{Fig:bump1}), we find
that this feature is $\sim 0.2$ mag fainter than in previous
identifications, but still fits the relation between the RGB-bump
absolute magnitude and the global metallicity of the system drawn by
Galactic GCs (see Fig.\ref{Fig:bump2}).

The King structural parameters are commonly used to estimate the
central and half-mass relaxation times of star clusters, by using
equations (10) and (11) in \citet{Djorgovski1993}. In practice,
however, this is not obvious in the case of core collapsed systems,
where the King model family is unable to properly reproduce the
observed density distribution. Keeping these limitations in mind, a
possible way to circumvent this problem is to estimate the central
relaxation time $t_{rc}$ from the King model that best fits the
innermost portion of the density profile (left panel in
Fig. \ref{fig:density_profile}), and the half-mass relaxation time
$t_{rh}$ from the King model reproducing the outer part of it (right
panel in the same figure). This is because $t_{rc}$ primarily depends
on the values of $r_c$, while $t_{rh}$ is mainly driven by the
half-mass radius $r_h$ \citep[see][]{Djorgovski1993}, and the two
mentioned profiles provide, respectively, the most appropriate
estimates of these two radial scales. Hence, by adopting
$r_c=2.8\arcsec$, together with $c=2.55$, the distance modulus and
reddening estimated in this work, and the integrated $V$-band
magnitude and central surface brightness quoted in the
\citet{Harris1996} catalogue, we find $\log(t_{rc})\simeq 6.72$ (in
units of years).  Adopting $r_h=43\arcsec$ (from the fit to the outer
portion of the profile), we find $\log(t_{rh})=9.06$. The latter value
is in good agreement with the estimates reported in
\citet{Harris1996} and in \citet{Baumgardt2018}, and we verified that
it remains almost unchanged even by adopting alternative King models
aimed at simultaneously reproducing as much as possible both the inner
and the outer portions of the density profile (see an example in
Figure \ref{fig:king_global}). This is because the value of $r_h$
varies only slightly.  Conversely, the central relaxation time quoted
in the \citet{Harris1996} catalogue is much larger, $\log(t_{rc})=7.24$,
mainly because of a larger value adopted for the core radius
($r_c=0.3$ pc), resulting from a fit to the surface brightness profile
obtained by fixing $c=2.5$ \citep{Trager1993}.  This highlights the
importance of forcing the fit to the innermost portion of the density
profile (as in the left panel of Fig. \ref{fig:density_profile}) if
one aims to estimate their central relaxation time of PCC clusters.
 
Table \ref{table:tab_params} summarises the value of the main
parameters obtained in this work for NGC 6284.  A solid determination
of the cluster distance from independent methods and the measure of
iron and $\alpha$-element abundances from high-resolution spectroscopy
are now urged to improve these values. In fact, once the distance and
the metallicity are properly fixed, the number of free parameters in
the isochrone fitting procedure would decrease and a more accurate
estimate of all the other quantities characterising this poorly
investigated cluster would be finally obtained.


\begin{acknowledgements}
This research is part of the project Cosmic-Lab ({\it "Star Clusters as cosmic laboratories"}) at the Physics and Astronomy Department of the University of Bologna (\url{http://www.cosmic-lab.eu/Cosmic-Lab/Home.html}). The research has been funded by project Light-on- Dark, granted by the Italian MIUR through contract PRIN-2017K7REXT (PI: Ferraro).
This work has made use of data from the European Space Agency (ESA) mission
{\it Gaia} (\url{https://www.cosmos.esa.int/gaia}), processed by the {\it Gaia}
Data Processing and Analysis Consortium (DPAC,
\url{https://www.cosmos.esa.int/web/gaia/dpac/consortium}). Funding for the DPAC
has been provided by national institutions, in particular the institutions
participating in the {\it Gaia} Multilateral Agreement.
\end{acknowledgements}

\bibliographystyle{aasjournal}

\begin{thebibliography}{}
    
\bibitem[{{Armandroff}(1989)}]{Armandroff1989}
{Armandroff}, T.~E. 1989, \aj, 97, 375.

\bibitem[Astropy Collaboration (2013)]{Astropy Collaboration} {Astropy Collaboration}, {Robitaille}, T.~P., {Tollerud}, E.~J., {et~al.} 2013, \aap, 558, A33. 

\bibitem[Astropy Collaboration, (2018)]{Astropy Collaboration} Astropy Collaboration, {Price-Whelan}, A.~M., {Sip{\H{o}}cz}, B.~M., {et~al.}, 2018, \aj, 156, 123.

\bibitem[{{Bajkova} \& {Bobylev}(2021)}]{Bajkova2021} {Bajkova}, A.~T., \& {Bobylev}, V.~V. 2021, Research in Astronomy and Astrophysics, 21, 173. 

\bibitem[Bhat et al.(2022)]{bhat+22} Bhat, B., Lanzoni, B., Ferraro, F.~R., et al.\ 2022, \apj, 926, 118. 

\bibitem[{{Baumgardt} \& {Hilker}(2018)}]{Baumgardt2018}
{Baumgardt}, H., \& {Hilker}, M. 2018, \mnras, 478, 1520.

\bibitem[{{Baumgardt} \& {Vasiliev}(2021)}]{Baumgardt2021}
{Baumgardt}, H., \& {Vasiliev}, E. 2021, \mnras, 505, 5957.

\bibitem[{{Bellazzini} {et~al.}(2020){Bellazzini}, {Ibata}, {Malhan}, {Martin},
  {Famaey}, \& {Thomas}}]{Bellazzini2020}
{Bellazzini}, M., {Ibata}, R., {Malhan}, K., {et~al.} 2020, \aap, 636, A107.

\bibitem[{{Bellini} {et~al.}(2011){Bellini}, {Anderson}, \&
  {Bedin}}]{Bellini2011}
{Bellini}, A., {Anderson}, J., \& {Bedin}, L.~R. 2011, \pasp, 123, 622.

\bibitem[Bellini et al.(2017)]{bellini+17} Bellini, A., Bianchini, P., Varri, A.~L., et al.\ 2017, \apj, 844, 167. 

\bibitem[{{Bertin} \& {Arnouts}(1996)}]{1996A&AS..117..393B}
{Bertin}, E., \& {Arnouts}, S. 1996, \aaps, 117, 393.

\bibitem[{{Bianchini} {et~al.}(2018){Bianchini}, {Webb}, {Sills}, \&
  {Vesperini}}]{Bianchini2018}
{Bianchini}, P., {Webb}, J.~J., {Sills}, A., \& {Vesperini}, E. 2018, \mnras,
  475, L96.

\bibitem[{{Cabrera-Ziri} \& {Conroy}(2022)}]{Cabrera2022}
{Cabrera-Ziri}, I., \& {Conroy}, C. 2022, \mnras, 511, 341.
 
\bibitem[Cadelano et~al.(2017a)]{Cadelano2017b}
{Cadelano}, M., {Dalessandro}, E., {Ferraro}, F.~R., {et~al.}
  2017{\natexlab{a}}, \apj, 836, 170.

\bibitem[{{Cadelano} {et~al.}(2022){Cadelano}, {Ferraro}, {Dalessandro},
  {Lanzoni}, {Pallanca}, \& {Saracino}}]{Cadelano2022}
{Cadelano}, M., {Ferraro}, F.~R., {Dalessandro}, E., {et~al.} 2022, \apj, 941,
  69.

\bibitem[{{Cadelano} {et~al.}(2019){Cadelano}, {Ferraro}, {Istrate},
  {Pallanca}, {Lanzoni}, \& {Freire}}]{Cadelano2019}
{Cadelano}, M., {Ferraro}, F.~R., {Istrate}, A.~G., {et~al.} 2019, \apj, 875,
  25.

\bibitem[{{Cadelano} {et~al.}(2017{\natexlab{b}}){Cadelano}, {Pallanca},
  {Ferraro}, {Dalessandro}, {Lanzoni}, \& {Patruno}}]{Cadelano2017a}
{Cadelano}, M., {Pallanca}, C., {Ferraro}, F.~R., {et~al.} 2017{\natexlab{b}},
  \apj, 844, 53.

\bibitem[{{Cadelano} {et~al.}(2018){Cadelano}, {Ransom}, {Freire}, {Ferraro},
  {Hessels}, {Lanzoni}, {Pallanca}, \& {Stairs}}]{Cadelano2018}
{Cadelano}, M., {Ransom}, S.~M., {Freire}, P.~C.~C., {et~al.} 2018, \apj, 855,
  125.

\bibitem[Cadelano et~al.(2020a)]{Cadelano2020a} {Cadelano}, M., {Saracino}, S., {Dalessandro}, E., {et~al.} 2020a, \apj, 895, 54.

\bibitem[Cadelano et~al. (2020b)]{Cadelano2020b}
{Cadelano}, M., {Chen}, J., {Pallanca}, C., {et~al.} 2020{\natexlab{b}}, \apj,
  905, 63.

 
\bibitem[{{Calamida} {et~al.}(2022){Calamida}, {Bajaj}, {Mack}, {Marinelli},
  {Medina}, {Pidgeon}, {Kozhurina-Platais}, {Shanahan}, \&
  {Som}}]{Calamida2022}
{Calamida}, A., {Bajaj}, V., {Mack}, J., {et~al.} 2022, \aj, 164, 32.

\bibitem[{{Cardelli} {et~al.}(1989){Cardelli}, {Clayton}, \&
  {Mathis}}]{Cardelli1989}
{Cardelli}, J.~A., {Clayton}, G.~C., \& {Mathis}, J.~S. 1989, \apj, 345, 245.

\bibitem[{{Carretta} {et~al.}(2009){Carretta}, {Bragaglia}, {Gratton},
  {D'Orazi}, \& {Lucatello}}]{Carretta2009}
{Carretta}, E., {Bragaglia}, A., {Gratton}, R., {D'Orazi}, V., \& {Lucatello},
  S. 2009, \aap, 508, 695.

\bibitem[{{Carretta} {et~al.}(2010){Carretta}, {Bragaglia}, {Gratton},
  {Recio-Blanco}, {Lucatello}, {D'Orazi}, \& {Cassisi}}]{Carretta2010}
{Carretta}, E., {Bragaglia}, A., {Gratton}, R.~G., {et~al.} 2010, \aap, 516,
  A55.

\bibitem[{{Casagrande} \& {VandenBerg}(2014)}]{Casagrande2014}
{Casagrande}, L., \& {VandenBerg}, D.~A. 2014, \mnras, 444, 392.

\bibitem[Casetti-Dinescu et al.(2010)]{casetti+10} Casetti-Dinescu, D.~I., Girard, T.~M., Korchagin, V.~I., et al.\ 2010, \aj, 140, 1282. 

\bibitem[{{Clement} {et~al.}(1980){Clement}, {Sawyer Hogg}, \&
  {Wells}}]{Clement1980}
{Clement}, C.~M., {Sawyer Hogg}, H., \& {Wells}, T.~R. 1980, \aj, 85, 1604.

\bibitem[Crociati et al.(2023)]{crociati+23} Crociati, C., Valenti, E., Ferraro, F.~R., et al.\ 2023, \apj, 951, 17. 

\bibitem[{{Dalessandro} {et~al.}(2019){Dalessandro}, {Ferraro}, {Bastian},
  {Cadelano}, {Lanzoni}, \& {Raso}}]{Dalessandro2019}
{Dalessandro}, E., {Ferraro}, F.~R., {Bastian}, N., {et~al.} 2019, \aap, 621,
  A45.

\bibitem[{{Dalessandro} {et~al.}(2013b){Dalessandro}, {Ferraro}, {Massari},
  {Lanzoni}, {Miocchi}, {Beccari}, {Bellini}, {Sills}, {Sigurdsson},
  {Mucciarelli}, \& {Lovisi}}]{Dalessandro2013}
{Dalessandro}, E., {Ferraro}, F.~R., {Massari}, D., {et~al.} 2013b, \apj, 778,
  135.

\bibitem[Dalessandro et al.(2013a)]{dalessandro+13a} Dalessandro, E., Salaris, M., Ferraro, F.~R., et al.\ 2013a, \mnras, 430, 459. 

\bibitem[Dalessandro et al.(2014)]{dalessandro+14} Dalessandro, E., Massari, D., Bellazzini, M., et al.\ 2014, \apjl, 791, L4. 

\bibitem[Dalessandro et al.(2022)]{dalessandro+22} Dalessandro, E., Crociati, C., Cignoni, M., et al.\ 2022, \apj, 940, 170. 

\bibitem[{{De Angeli} {et~al.}(2005){De Angeli}, {Piotto}, {Cassisi}, {Busso},
  {Recio-Blanco}, {Salaris}, {Aparicio}, \& {Rosenberg}}]{Deangeli2005}
{De Angeli}, F., {Piotto}, G., {Cassisi}, S., {et~al.} 2005, \aj, 130, 116.

\bibitem[{{Deras} {et~al.}(2023){Deras}, {Cadelano}, {Ferraro}, {Lanzoni}, \&
  {Pallanca}}]{Deras2023}
{Deras}, D., {Cadelano}, M., {Ferraro}, F.~R., {Lanzoni}, B., \& {Pallanca}, C.
  2023, \apj, 942, 104.

\bibitem[{{Dias} {et~al.}(2016){Dias}, {Barbuy}, {Saviane}, {Held}, {Da Costa},
  {Ortolani}, {Gullieuszik}, \& {V{\'a}squez}}]{Dias2016}
{Dias}, B., {Barbuy}, B., {Saviane}, I., {et~al.} 2016, \aap, 590, A9.

\bibitem[{{Djorgovski}(1993)}]{Djorgovski1993}
{Djorgovski}, S. 1993, in Astronomical Society of the Pacific Conference
  Series, Vol.~50, Structure and Dynamics of Globular Clusters, ed. S.~G.
  {Djorgovski} \& G.~{Meylan}, 373

\bibitem[Djorgovski \& King(1986)]{djo+86} Djorgovski, S. \& King, I.~R.\ 1986, \apjl, 305, L61. 

\bibitem[{{Dotter} {et~al.}(2008){Dotter}, {Chaboyer},  {Jevremovi{\'c}}, {Kostov}, {Baron}, \& {Ferguson}}]{Dotter2008}  {Dotter}, A., {Chaboyer}, B., {Jevremovi{\'c}}, D., {et~al.} 2008,
  \apjs, 178, 89.

\bibitem[Ferraro et al.(1991)]{ferraro+91} Ferraro, F.~R., Clementini, G., Fusi Pecci, F., et al.\ 1991, \mnras, 252, 357. 

\bibitem[Ferraro et al.(1992)]{ferraro+92} Ferraro, F.~R., Clementini, G., Fusi Pecci, F., et al.\ 1992, \mnras, 256, 391. 

\bibitem[Ferraro et al.(1997)]{ferraro+97} Ferraro, F.~R., Paltrinieri, B., Fusi Pecci, F., et al.\ 1997, \apjl, 484, L145. 

\bibitem[Ferraro et al.(1999a)]{ferraro+99a} Ferraro, F.~R., Messineo, M., Fusi Pecci, F., et al.\ 1999a, \aj, 118, 1738. 

\bibitem[Ferraro et al.(1999b)]{ferraro+99b} Ferraro, F.~R., Paltrinieri, B., Rood, R.~T., et al.\ 1999b, \apj, 522, 983. 

\bibitem[Ferraro et al.(2000)]{ferraro+00} Ferraro, F.~R., Montegriffo, P., Origlia, L., et al.\ 2000, \aj, 119, 1282. 
 
\bibitem[Ferraro et~al.(2003)]{Ferraro2003}
{Ferraro}, F.~R., {Possenti}, A., {Sabbi}, E., {et~al.} 2003, \apj, 595, 179.

\bibitem[Ferraro et~al.(2009a)]{Ferraro2009}
{Ferraro}, F.~R., {Dalessandro}, E., {Mucciarelli}, A., {et~al.}
  2009{\natexlab{a}}, \nat, 462, 483.

\bibitem[Ferraro et~al.(2009b)]{Ferraro2009b}
{Ferraro}, F.~R., {Beccari}, G., {Dalessandro}, E., {et~al.}
  2009{\natexlab{b}}, \nat, 462, 1028.

\bibitem[Ferraro et al.(2012)]{ferraro+12} Ferraro, F.~R., Lanzoni, B., Dalessandro, E., et al.\ 2012, \nat, 492, 393. 

\bibitem[Ferraro et~al.(2016)]{Ferraro2016}
{Ferraro}, F.~R., {Massari}, D., {Dalessandro}, E., {et~al.} 2016, \apj, 828,
  75.
 
\bibitem[Ferraro et al.(2018a)]{ferraro+18a} Ferraro, F.~R., Mucciarelli, A., Lanzoni, B., et al.\ 2018a, \apj, 860, 50. 

\bibitem[Ferraro et al.(2018b)]{ferraro+18b} Ferraro, F.~R., Lanzoni, B., Raso, S., et al.\ 2018b, \apj, 860, 36. 

\bibitem[Ferraro et al.(2019)]{ferraro+19} Ferraro, F.~R., Lanzoni, B., Dalessandro, E., et al.\ 2019, Nature Astronomy, 3, 1149. 

\bibitem[Ferraro et~al.(2021)]{Ferraro2021} {Ferraro}, F.~R., {Pallanca}, C., {Lanzoni}, B., {et~al.} 2021, Nature Astronomy, 5, 311.

\bibitem[Ferraro et al.(2023)]{ferraro+23} Ferraro, F.~R., Lanzoni, B., Vesperini, E., et al.\ 2023, \apj, 950, 145. 

\bibitem[{{Foreman-Mackey} {et~al.}(2013){Foreman-Mackey}, {Hogg}, {Lang}, \&
  {Goodman}}]{Foreman2013}
{Foreman-Mackey}, D., {Hogg}, D.~W., {Lang}, D., \& {Goodman}, J. 2013, \pasp,
  125, 306.

\bibitem[{{Foreman-Mackey} {et~al.}(2019){Foreman-Mackey}, {Farr}, {Sinha},
  {Archibald}, {Hogg}, {Sanders}, {Zuntz}, {Williams}, {Nelson}, {de
  Val-Borro}, {Erhardt}, {Pashchenko}, \& {Pla}}]{Foreman2019}
{Foreman-Mackey}, D., {Farr}, W., {Sinha}, M., {et~al.} 2019, The Journal of
  Open Source Software, 4, 1864.

\bibitem[Fusi Pecci et al.(1990)]{fusi+90} Fusi Pecci, F., Ferraro, F.~R., Crocker, D.~A., et al.\ 1990, \aap, 238, 95

\bibitem[Gaia Collaboration et~al.(2022)]{Gaia2022}
{Gaia Collaboration}, {Vallenari}, A., {Brown}, A.~G.~A., {et~al.} 2022, arXiv
  e-prints, arXiv:2208.00211.

\bibitem[{{Girardi} {et~al.}(2002){Girardi}, {Bertelli}, {Bressan}, {Chiosi},
  {Groenewegen}, {Marigo}, {Salasnich}, \& {Weiss}}]{Girardi2002}
{Girardi}, L., {Bertelli}, G., {Bressan}, A., {et~al.} 2002, \aap, 391, 195.

\bibitem[{{Gnedin} {et~al.}(1999){Gnedin}, {Lee}, \& {Ostriker}}]{Gnedin1999}
{Gnedin}, O.~Y., {Lee}, H.~M., \& {Ostriker}, J.~P. 1999, \apj, 522, 935.

\bibitem[{{Gnedin} \& {Ostriker}(1997)}]{Gnedin1997}
{Gnedin}, O.~Y., \& {Ostriker}, J.~P. 1997, \apj, 474, 223.

\bibitem[{{Harris}(1996)}]{Harris1996}
{Harris}, W.~E. 1996, \aj, 112, 1487

\bibitem[Harris(2018)]{harris+18} Harris, W.~E.\ 2018, \aj, 156, 296. 

\bibitem[{{H{\'e}non}(1961)}]{Henon1961}
{H{\'e}non}, M. 1961, Annales d'Astrophysique, 24, 369

\bibitem[{{Hills} \& {Day}(1976)}]{Hills1976}
{Hills}, J.~G., \& {Day}, C.~A. 1976, \aplett, 17, 87

\bibitem[Kamann et al.(2018)]{kamann+18} Kamann, S., Husser, T.-O., Dreizler, S., et al.\ 2018, \mnras, 473, 5591. 

\bibitem[{{King}(1966)}]{King1966}
{King}, I.~R. 1966, \aj, 71, 64.

\bibitem[{{Kundu} {et~al.}(2022){Kundu}, {Navarrete}, {Sbordone},
  {Carballo-Bello}, {Fern{\'a}ndez-Trincado}, {Minniti}, \&
  {Singh}}]{Kundu2022}
{Kundu}, R., {Navarrete}, C., {Sbordone}, L., {et~al.} 2022, \aap, 665, A8.

\bibitem[Ibata et al.(2009)]{ibata+09} Ibata, R., Bellazzini, M., Chapman, S.~C., et al.\ 2009, \apjl, 699, L169. 

\bibitem[Lanzoni et~al.(2007a)]{Lanzoni2007a}
{Lanzoni}, B., {Dalessandro}, E., {Ferraro}, F.~R., {et~al.}
  2007{\natexlab{a}}, \apj, 663, 267.

\bibitem[Lanzoni et~al.(2007b)]{Lanzoni2007b} Lanzoni et al., 2007b, \apjl, 668, L139.

\bibitem[Lanzoni et~al.(2007c)]{Lanzoni2007c} {Lanzoni}, B., {Sanna}, N., {Ferraro}, F.~R., {et~al.} 2007{\natexlab{c}}, \apj, 663, 1040.

\bibitem[Lanzoni et~al.(2010)]{Lanzoni2010}
{Lanzoni}, B., {Ferraro}, F.~R., {Dalessandro}, E., {et~al.} 2010, \apj, 717,
  653.
 
\bibitem[Lanzoni et al.(2016)]{lanzoni+16} Lanzoni, B., Ferraro, F.~R., Alessandrini, E., et al.\ 2016, \apjl, 833, L29. 

\bibitem[Lanzoni et al.(2018a)]{lanzoni+18a} Lanzoni, B., Ferraro, F.~R., Mucciarelli, A., et al.\ 2018a, \apj, 861, 16. 

\bibitem[Lanzoni et al.(2018b)]{lanzoni+18b} Lanzoni, B., Ferraro, F.~R., Mucciarelli, A., et al.\ 2018b, \apj, 865, 11.  

\bibitem[Lanzoni et~al.(2019)]{Lanzoni2019} Lanzoni et al., 
 2019, \apj, 887, 176.

\bibitem[Leanza et al.(2022)]{leanza+22} Leanza, S., Pallanca, C., Ferraro, F.~R., et al.\ 2022, \apj, 929, 186. 

\bibitem[Leanza et al.(2023)]{leanza+23} Leanza, S., Pallanca, C., Ferraro, F.~R., et al.\ 2023, \apj, 944, 162. 

\bibitem[{{Lynden-Bell} \& {Wood}(1968)}]{LyndenBell1968}
{Lynden-Bell}, D., \& {Wood}, R. 1968, \mnras, 138, 495.

\bibitem[Lugger et al.(1995)]{lugger+95} Lugger, P.~M., Cohn, H.~N., \& Grindlay, J.~E.\ 1995, \apj, 439, 191. 

\bibitem[{{Marigo} {et~al.}(2017){Marigo}, {Girardi}, {Bressan}, {Rosenfield},
  {Aringer}, {Chen}, {Dussin}, {Nanni}, {Pastorelli}, {Rodrigues}, {Trabucchi},
  {Bladh}, {Dalcanton}, {Groenewegen}, {Montalb{\'a}n}, \& {Wood}}]{Marigo2017}
{Marigo}, P., {Girardi}, L., {Bressan}, A., {et~al.} 2017, \apj, 835, 77.

\bibitem[{{Massari} {et~al.}(2019){Massari}, {Koppelman}, \&
  {Helmi}}]{Massari2019}
{Massari}, D., {Koppelman}, H.~H., \& {Helmi}, A. 2019, \aap, 630, L4.

\bibitem[{{Massari} {et~al.}(2014){Massari}, {Mucciarelli}, {Ferraro},
  {Origlia}, {Rich}, {Lanzoni}, {Dalessandro}, {Valenti}, {Ibata}, {Lovisi},
  {Bellazzini}, \& {Reitzel}}]{Massari2014}
{Massari}, D., {Mucciarelli}, A., {Ferraro}, F.~R., {et~al.} 2014, \apj, 795,
  22.

\bibitem[{{McCrea}(1964)}]{McCrea1964}
{McCrea}, W.~H. 1964, \mnras, 128, 147.

\bibitem[{{Meissner} \& {Weiss}(2006)}]{Meissner2006}
{Meissner}, F., \& {Weiss}, A. 2006, \aap, 456, 1085.

\bibitem[{{Meylan} \& {Heggie}(1997)}]{Meylan1997}
{Meylan}, G., \& {Heggie}, D.~C. 1997, \aapr, 8, 1.

\bibitem[{{Minniti} {et~al.}(1995){Minniti}, {Olszewski}, \&
  {Rieke}}]{Minniti1995}
{Minniti}, D., {Olszewski}, E.~W., \& {Rieke}, M. 1995, \aj, 110, 1686.

\bibitem[{{Miocchi} {et~al.}(2013){Miocchi}, {Lanzoni}, {Ferraro},
  {Dalessandro}, {Vesperini}, {Pasquato}, {Beccari}, {Pallanca}, \&
  {Sanna}}]{Miocchi2013}
{Miocchi}, P., {Lanzoni}, B., {Ferraro}, F.~R., {et~al.} 2013, \apj, 774, 151.

\bibitem[{{Moffat}(1969)}]{Moffat1969}
{Moffat}, A.~F.~J. 1969, \aap, 3, 455

\bibitem[{{Montegriffo} {et~al.}(1995){Montegriffo}, {Ferraro}, {Fusi Pecci},
  \& {Origlia}}]{Montegriffo1995}
{Montegriffo}, P., {Ferraro}, F.~R., {Fusi Pecci}, F., \& {Origlia}, L. 1995,
  \mnras, 276, 739.

\bibitem[Nataf et al.(2013)]{nataf+13} Nataf, D.~M., Gould, A.~P., Pinsonneault, M.~H., et al.\ 2013, \apj, 766, 77. 

\bibitem[{{Noyola} \& {Gebhardt}(2006)}]{Noyola2006}
{Noyola}, E., \& {Gebhardt}, K. 2006, \aj, 132, 447.

\bibitem[Origlia et al.(2002)]{origlia+02} Origlia, L., Ferraro, F.~R., Fusi Pecci, F., et al.\ 2002, \apj, 571, 458. 

\bibitem[Origlia et al.(2003)]{origlia+03} Origlia, L., Ferraro, F.~R., Bellazzini, M., et al.\ 2003, \apj, 591, 916. 

\bibitem[{{Origlia} {et~al.}(2013){Origlia}, {Massari}, {Rich}, {Mucciarelli},
  {Ferraro}, {Dalessandro}, \& {Lanzoni}}]{Origlia2013}
{Origlia}, L., {Massari}, D., {Rich}, R.~M., {et~al.} 2013, \apjl, 779, L5.

\bibitem[{{Origlia} {et~al.}(2011){Origlia}, {Rich}, {Ferraro}, {Lanzoni},
  {Bellazzini}, {Dalessandro}, {Mucciarelli}, {Valenti}, \&
  {Beccari}}]{Origlia2011}
{Origlia}, L., {Rich}, R.~M., {Ferraro}, F.~R., {et~al.} 2011, \apjl, 726, L20.

\bibitem[{{Pallanca} {et~al.}(2019){Pallanca}, {Ferraro}, {Lanzoni},
  {Saracino}, {Raso}, \& {Focardi}}]{Pallanca2019}
{Pallanca}, C., {Ferraro}, F.~R., {Lanzoni}, B., {et~al.} 2019, \apj, 882, 159.

\bibitem[Pallanca et~al.(2021a)]{Pallanca2021a}
{Pallanca}, C., {Ferraro}, F.~R., {Lanzoni}, B., {et~al.} 2021a,
  \apj, 917, 92.


\bibitem[Pallanca et~al.(2021b)]{Pallanca2021b}
{Pallanca}, C., {Lanzoni}, B., {Ferraro}, F.~R., {et~al.} 2021b,
  \apj, 913, 137.
 
\bibitem[Pallanca et al.(2023)]{pallanca+23} Pallanca, C., Leanza, S., Ferraro, F.~R., et al.\ 2023, \apj, 950, 138. 

\bibitem[{{Penny}(1976)}]{Penny1976}
{Penny}, A.~J. 1976, {Electronographic stellar photometry}

\bibitem[{{Pietrinferni} {et~al.}(2021){Pietrinferni}, {Hidalgo}, {Cassisi},
  {Salaris}, {Savino}, {Mucciarelli}, {Verma}, {Silva Aguirre}, {Aparicio}, \&
  {Ferguson}}]{Pietrinferni2021}
{Pietrinferni}, A., {Hidalgo}, S., {Cassisi}, S., {et~al.} 2021, \apj, 908,
  102.

\bibitem[{{Piotto} {et~al.}(2002){Piotto}, {King}, {Djorgovski}, {Sosin},
  {Zoccali}, {Saviane}, {De Angeli}, {Riello}, {Recio-Blanco}, {Rich},
  {Meylan}, \& {Renzini}}]{Piotto2002}
{Piotto}, G., {King}, I.~R., {Djorgovski}, S.~G., {et~al.} 2002, \aap, 391,
  945.

\bibitem[{{Raso} {et~al.}(2020){Raso}, {Libralato}, {Bellini}, {Ferraro},
  {Lanzoni}, {Cadelano}, {Pallanca}, {Dalessandro}, {Piotto}, {Anderson}, \&
  {Sohn}}]{Raso2020}
{Raso}, S., {Libralato}, M., {Bellini}, A., {et~al.} 2020, \apj, 895, 15.

\bibitem[{{Recio-Blanco} {et~al.}(2005){Recio-Blanco}, {Piotto}, {de Angeli},
  {Cassisi}, {Riello}, {Salaris}, {Pietrinferni}, {Zoccali}, \&
  {Aparicio}}]{Recio_Blanco2005}
{Recio-Blanco}, A., {Piotto}, G., {de Angeli}, F., {et~al.} 2005, \aap, 432,
  851.

\bibitem[{{Reed} {et~al.}(1988){Reed}, {Hesser}, \& {Shawl}}]{Reed1988}
{Reed}, B.~C., {Hesser}, J.~E., \& {Shawl}, S.~J. 1988, \pasp, 100, 545.

\bibitem[Romano et al.(2023)]{romano+23} Romano, D., Ferraro, F.~R., Origlia, L., et al.\ 2023, \apj, 951, 85. 

\bibitem[Salaris et al.(1993)]{salaris+93} Salaris, M., Chieffi, A., \& Straniero, O.\ 1993, \apj, 414, 580. 

\bibitem[Salaris \& Cassisi(2006)]{salariscassisi} Salaris, M. \&
  Cassisi, S.\ 2006, Evolution of Stars and Stellar Populations by
  Maurizio Salaris and Santi Cassisi. Wiley, 2006. ISBN:
  978-0-470-09219-4

\bibitem[{{Saracino} {et~al.}(2016){Saracino}, {Dalessandro}, {Ferraro},
  {Geisler}, {Mauro}, {Lanzoni}, {Origlia}, {Miocchi}, {Cohen}, {Villanova}, \&
  {Moni Bidin}}]{Saracino2016}
{Saracino}, S., {Dalessandro}, E., {Ferraro}, F.~R., {et~al.} 2016, \apj, 832,
  48.

\bibitem[{{Saracino} {et~al.}(2019){Saracino}, {Dalessandro}, {Ferraro},
  {Lanzoni}, {Geisler}, {Cohen}, {Bellini}, {Vesperini}, {Salaris}, {Cassisi},
  {Pietrinferni}, {Origlia}, {Mauro}, {Villanova}, \& {Moni
  Bidin}}]{Saracino2019}
---. 2019, \apj, 874, 86.

\bibitem[{{Simunovic} {et~al.}(2014){Simunovic}, {Puzia}, \&
  {Sills}}]{Simunovic2014}
{Simunovic}, M., {Puzia}, T.~H., \& {Sills}, A. 2014, \apjl, 795, L10.

\bibitem[{{Smith} \& {Perkins}(1982)}]{Smith1982}
{Smith}, H.~A., \& {Perkins}, G.~J. 1982, \apj, 261, 576.

\bibitem[{{Stetson}(1987)}]{Stetson1987}
{Stetson}, P.~B. 1987, \pasp, 99, 191.

\bibitem[{{Stetson}(1994)}]{Stetson1994}
---. 1994, \pasp, 106, 250.

\bibitem[{{Trager} {et~al.}(1993){Trager}, {Djorgovski}, \&
  {King}}]{Trager1993}
{Trager}, S.~C., {Djorgovski}, S., \& {King}, I.~R. 1993, in Astronomical
  Society of the Pacific Conference Series, Vol.~50, Structure and Dynamics of
  Globular Clusters, ed. S.~G. {Djorgovski} \& G.~{Meylan}, 347

\bibitem[{{Trager} {et~al.}(1995){Trager}, {King}, \&
  {Djorgovski}}]{Trager1995}
{Trager}, S.~C., {King}, I.~R., \& {Djorgovski}, S. 1995, \aj, 109, 218.

\bibitem[{{Usher} {et~al.}(2019){Usher}, {Beckwith}, {Bellstedt}, {Alabi},
  {Chevalier}, {Pastorello}, {Cerulo}, {Dalgleish}, {Fraser-McKelvie},
  {Kamann}, {Penny}, {Foster}, {McDermid}, {Schiavon}, \&
  {Villaume}}]{Usher2019}
{Usher}, C., {Beckwith}, T., {Bellstedt}, S., {et~al.} 2019, \mnras, 482, 1275.

\bibitem[Valenti et al.(2004)]{valenti+04} Valenti, E., Ferraro, F.~R., \& Origlia, L.\ 2004, \mnras, 354, 815. 

\bibitem[{{Valenti} {et~al.}(2007){Valenti}, {Ferraro}, \&
  {Origlia}}]{Valenti2007}
{Valenti}, E., {Ferraro}, F.~R., \& {Origlia}, L. 2007, \aj, 133, 1287.

\bibitem[Valenti et~al.(2010)]{Valenti2010} Valenti E., {Ferraro}, F.~R., \& {Origlia}, L. 2010, \mnras, 402, 1729.

\bibitem[{{Webbink}(1985)}]{Webbink1985}
{Webbink}, R.~F. 1985, in Dynamics of Star Clusters, ed. J.~{Goodman} \&
  P.~{Hut}, Vol. 113, 541--577

\bibitem[{{Zinn}(1985)}]{Zinn1985}
{Zinn}, R. 1985, \apj, 293, 424.

\bibitem[Zinn \& West(1984)]{zinnwest84} Zinn, R. \& West, M.~J.\ 1984, \apjs, 55, 45. 
  
\bibitem[Zoccali et al.(1999)]{zoccali+99} Zoccali, M., Cassisi, S., Piotto, G., et al.\ 1999, \apjl, 518, L49. 

\bibitem[{{Zocchi} {et~al.}(2016){Zocchi}, {Gieles}, {H{\'e}nault-Brunet}, \&
  {Varri}}]{Zocchi2016}
{Zocchi}, A., {Gieles}, M., {H{\'e}nault-Brunet}, V., \& {Varri}, A.~L. 2016,
  \mnras, 462, 696.



\end{thebibliography}



\end{document}